\documentclass{natastro}

\articletype{Article}
\articledoi{10.1038/s41550-026-03025-y.}       
\preprintnote{Published in \textit{Nature Astronomy}}

\usepackage{comment}

\title{Excess intergalactic heating around a dual quasar system 2.8 Gyr after the Big Bang}

\author{Kirsten R. Hall\textsuperscript{1}\thanks{Equal contribution.}
  \and Nadia L. Zakamska\textsuperscript{2}
  \and Jake S. Bennett\textsuperscript{3}
  \and Yu-Ching Chen \textsuperscript{2}
  \and Xiaohui Fan \textsuperscript{4}
  }
\affiliation{\textsuperscript{1}Radio and Geoastronomy Division, Center for Astrophysics $\vert$ Harvard \& Smithsonian, 60 Garden St., Cambridge, 02138, MA, USA
  \textsuperscript{2}The William H. Miller III Department of Physics and Astronomy, Johns Hopkins University, 3701 San Martin Drive, Baltimore, 21218, MD, USA
  \textsuperscript{3}Institute for Theory and Computation, Center for Astrophysics $\vert$ Harvard \& Smithsonian, 60 Garden St., Cambridge, 02138, MA, USA
  \textsuperscript{4}Steward Observatory, University of Arizona, 933 N Cherry Avenue, Tucson, 85721, Z, USA}
\correspondence{kirsten.hall.cfa.harvard.edu}

\begin{document}

\begin{abstract}
Galaxy formation models necessitate that actively growing supermassive black holes, or quasars, strongly impact their environments. Theory suggests that quasar winds produce very hot gas bubbles that extend through the host galaxy and into the circumgalactic medium. Because this gas component is so hot and diffuse, the only way to measure its total energy is through the thermal Sunyaev-Zel’dovich (tSZ) effect.  Here we report the discovery of a dual quasar system at z=2.37 which is embedded in a bubble of hot gas as detected (at 6$\sigma$) by a tSZ flux decrement at millimeter wavelengths. The total thermal energy of the hot gas is 9.7$\times$10$^{60}$~ergs. This energy exceeds the amount expected from all of the hot gas in a 10$^{13}$~M$_\odot$ dark matter halo by a factor of 3.6, and constitutes at least 0.6\% of the quasar’s lifetime energy output. We also find evidence for hot gas on larger (100-600~kpc) scales that may also originate in quasar outflows. This considerable excess energy is more than sufficient to shut off star formation and heat the intergalactic medium, implying a significant impact from quasar-driven winds on their surroundings. Such feedback plausibly explains the steep decline of the bright end of the luminosity function of galaxies across the observable Universe, and thus provides evidence to further support leading galaxy evolution theory. 
\end{abstract}

\maketitle

The observed tight relationship between the masses of black holes in galaxies and the velocities and masses of their stellar populations \cite{mago98, gebh00, ferr00} has made it clear that the active (quasar) phase of black hole evolution has profound effects on the formation of the galaxy. 
In the last five to ten years, some of the most compelling, direct evidence for feedback from quasar-driven outflows has been uncovered using the thermal Sunyaev-Zel'dovich effect (\cite{ruan15, verd16, cric16, soer17, lacy19, hall19, brow19}). 
The thermal Sunyaev-Zel'dovich (tSZ) effect  \cite{suny70}  is the inverse Compton scattering of Cosmic Microwave Background (CMB) photons off of free electrons in an intervening hot gas, such as in the virialized haloes hosting galaxy clusters and in gas bubbles that are ubiquitously theorized to exist due to quasar-driven outflows \cite{fauc12,zubo12,gill17}.
There is new observational evidence that baryonic feedback is more extended and more powerful than previously thought \cite{hadz25a, pand25, hadz25b, ried25}.
We are now entering the era of spatially resolving the tSZ effect with large, high-resolution millimeter telescopes and interferometers \cite{mroc19}. 
The increased sensitivity of these instruments allows measurements of the tSZ down to lower halo masses, and consequently, to higher redshifts \cite{goba19, dima23}. 

Importantly, the tSZ does not suffer from cosmological dimming; the surface brightness does not decrease as a function of distance to the intervening gas because the CMB photons enter from all directions \cite{suny70, suny72}.
The tSZ effect is a particularly powerful tool for probing gas that is very hot (T$>$10$^{8}$ K) and diffuse, and thus difficult to detect by emission line diagnostics at redshifts around cosmic noon.
Obtaining robust measurements of the thermal medium at these temperatures takes hundreds of kiloseconds of observing time with the best X-ray facilities, and even then it is difficult to separate the emission from the thermal and non-thermal components when a quasar is present \cite{cari98,over05,cham21,tozz22}.
Further, because the probability of scattering depends linearly on the product of temperature and density, the signal from the tSZ effect provides ground-based, nearly redshift-independent tool to probe the total thermal energy of the hot gas \cite{suny70}. 

Thus far, the majority of the detections of the tSZ effect in quasar environments are through cross-correlations of tens to hundreds of thousands of quasars with millimeter maps covering thousands of square degrees on the sky \cite{ruan15, verd16, cric16, soer17, hall19}, with only the exception of \cite{lacy19} and \cite{brow19}, and tentatively \cite{jone23}. 
The fractional energy deposition with respect to the quasar luminosity output in the stacking studies (e.g., \cite{hall19}) are $\sim$5\%, but for individual sources it is quite low ($\leq$0.01\%).
Simulations are often tuned to reproduce the observed galaxy luminosity function by either heating or expelling the gas to larger radii to prevent more stars from forming in a process known as negative feedback. 
Both scenarios require a minimum coupling fraction that is at least a few percent of the quasar energy output, and there is significant uncertainty around which version of feedback (heating vs. expulsion) is the dominant effect \cite{Benn25}.
Thus, empirical measurements of the tSZ effect around individual objects are crucial diagnostics for understanding how feedback transpires in quasar host galaxies and their environments.

We used multiwavelength observations of the hyperluminous (L$_\mathrm{bol}$ = 10$^{47.7}$ erg/s) quasar J154938.71+124509.1 ($z=2.374$) and its environment obtained with the Karl G. Jansky Very Large Array (VLA), the Atacama Large Millimeter/Submillimeter Array (ALMA), and the Submillimeter Array (SMA). 
We examine the radio through sub-mm spectral energy distributions (SEDs) of the radio-quiet quasar with a dust-obscured quasar companion in order to physically characterize their emission and to detect the tSZ effect due to the hottest gas phase of quasar winds.
We also report as supplementary information the measurement of two new energy transitions of CO emission from both sources, which support our conclusions. We use a $\Lambda$CDM cosmology with H$_0 =$67.4~km~s$^{-1}$~Mpc$^{-1}$ and $\Omega_m$ = 0.315 \cite{plan20}.

\section{Results}\label{sec2}

\begin{figure}[!htbp]
\centering
	\includegraphics[trim=0cm 0cm 1cm 0cm, clip, width=\columnwidth]{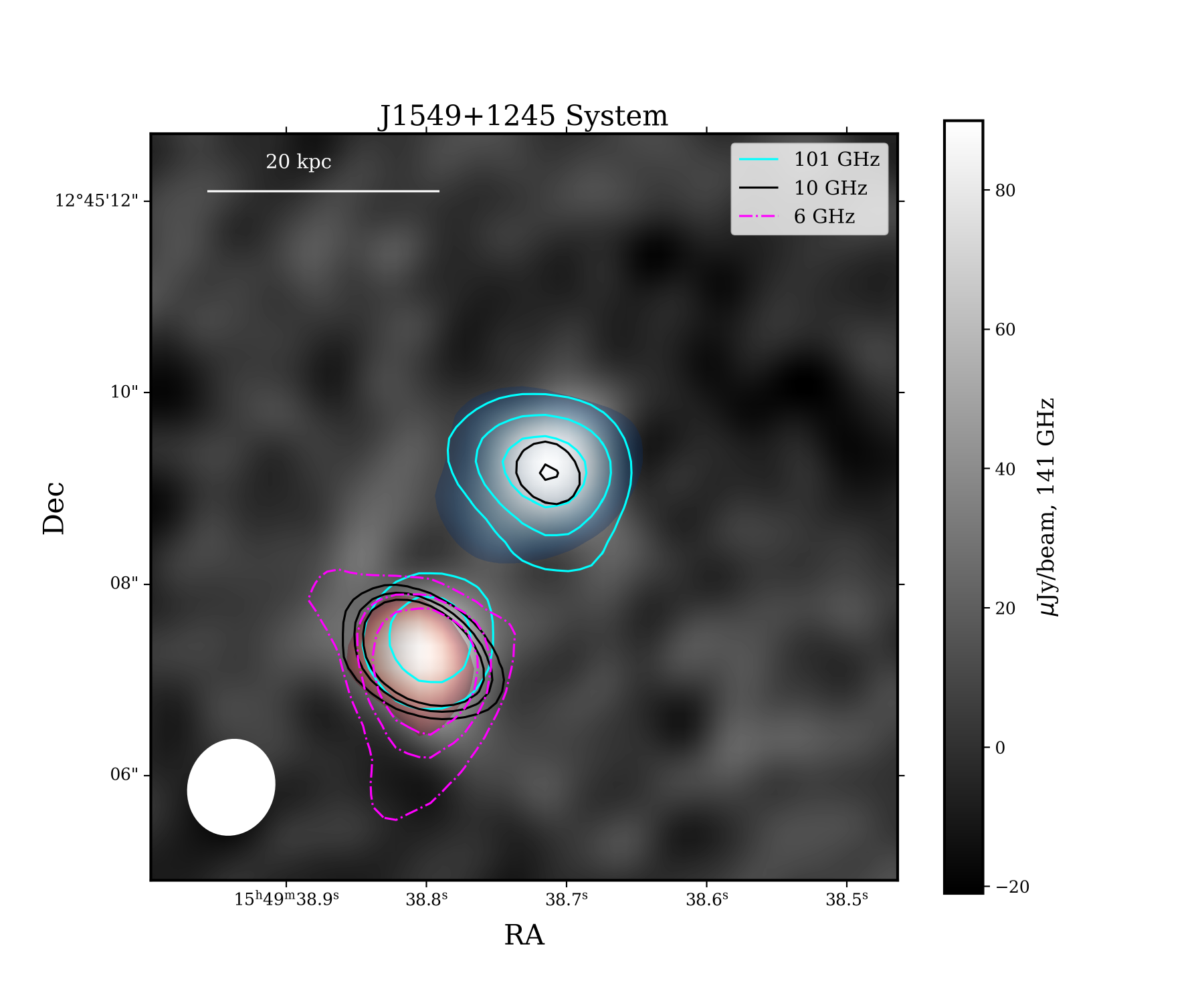}
    \captionof{figure}{Multi-frequency image of the J1549+1245 system. The background shading is a 141~GHz image, centered on J1549A and J1549B with 6~GHz (magenta dot-dashed, 1.6"x1"), 10~GHz (black, 1"x0.7"),  101~GHz (cyan, 1.3"x1.2"), and the CO(3-2) emission overlaid in blue and red. Contours are 4, 7, and 10$\sigma$, where $\sigma$= 36, 11, and 5.3 $\mu$Jy for the 6, 10, and 97~GHz images, respectively. The 141~GHz map has an rms $\sigma$=7$\mu$Jy and a beam size of 1.0"x0.8" (lower left ellipse).}
    \label{fig:J1549im}
\end{figure}

\subsection{J1549+1245 System Spectral Energy Distributions}

Our target, J154938.71+124509.1 (J1549A from here on) is from the WISE-SDSS selected hyper-luminous (WISSH, \cite{bisc17}) quasar sample. 
It has an optically undetected companion (J1549B) which is seen in radio and sub-mm, as well in multiple emission lines of CO. J1549B is just 19~kpc distant (950~km/s spectrally separated) from J1549A (Figure~\ref{fig:J1549im}) \cite{bisc17,bisc21,viet20}. 


We extract the SEDs from 0.8-2" apertures, thus probing the physical processes on galaxy-wide ($\sim$7-17~kpc) scales.
The physical nature of the companion galaxy J1549B is revealed in its SED of emission in the near-IR through radio. 
J1549B is a massive (M$_\mathrm{dyn} \approx$ 3.6$\times$10$^{11}$~M$_\odot$ \cite{bisc21}) dusty (L$_{IR}\sim$10$^{12}$L$_\odot$, undetected in optical/near-IR observations) galaxy  that likely hosts a powerful quasar as indicated by the relative strength of the radio and near-infrared fluxes in comparison to the far-IR and submillimeter emission (Supplementary Section~1, Supplementary Figure~1). 
Star formation in the host galaxy cannot plausibly produce the observed levels of radio and near-IR emission (Supplementary Section~1 and  \cite{bell03}).

In radio through sub-mm, J1549A is not detected below 10~GHz or above 233~GHz, but is detected at these two frequencies, as well as at 34, 101, and 145~GHz.
The flux densities of J1549A are fit with a combined SED made up of a power law (likely from free-free emission) and the modified black body describing the cold dust emission (Equation~\ref{eq:Sdust}), the details of which are shown in the Supplementary, Section~1 (Supplementary Figure~2). 

\begin{figure*}[!tbp]
\centering
       \includegraphics[width=\textwidth]{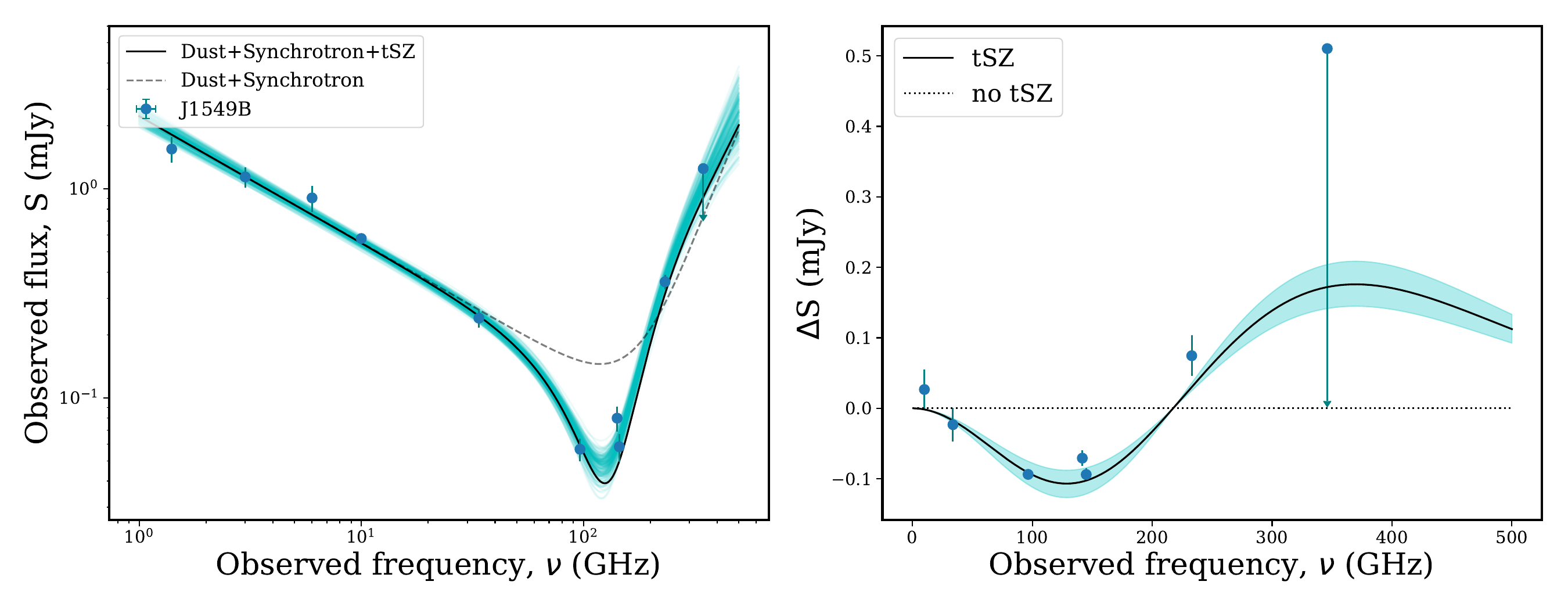}
    \captionof{figure}{Spectral energy distribution and modeling of J1549B. Left: Flux points are extracted from the images and uncertainties are the quadrature sum of the root-mean-square uncertainty and the flux calibration uncertainty. The black line is the model generated from the medians of the posterior parameter distributions describing the synchrotron emission, thermal Sunyaev-Zel'dovich effect spectral distortion, and the modified blackbody describing the cold dust emission. The cyan lines are models generated from 200 random draws within the 16th and 84th percentiles of posterior distributions of each parameter. J1549B's 345~GHz flux is an upper limit, measured at the level of 2.5$\sigma$, and not included in the fit. The dashed line depicts the model fit by a spectrum of synchrotron and modified blackbody only. Right: Residual fluxes and associated uncertainties from 10-345~GHz after subtracting away the model fit by a spectrum of synchrotron plus modified blackbody only. The black curve is the resulting thermal Sunyaev-Zel'dovich spectrum using the median of the posterior parameter distribution for the amplitude of the spectral distortion. The shading depicts the uncertainty as determined by the 16th and 84th percentiles of this parameter distribution. The dotted line is the expected residuals if there were no spectral distortion from the tSZ in the measured spectral energy distribution.}
    \label{fig:J1549SED}
\end{figure*}

J1549B is detected in 9 observations from 1.4~GHz to 233~GHz and has an upper limit (2.5$\sigma$ measurement) in the SMA data at 345~GHz. 
We model the SED of J1549B using the addition of a power law synchrotron spectrum, a thermal dust spectrum (Equation~\ref{eq:Sdust}), and the tSZ spectral distortion (Equation~\ref{eq:Ssz}). 
Understanding the radio emission mechanisms in radio quiet quasars remains a topic of debate \cite{zaka16, hwan18, laor19, pane19}. 
Adding to the limited sample of radio observations in such objects, especially at $z>2$ for which we can fully characterize the radio spectral shapes is of independent value in this study.
The medians of the posterior parameter distributions (which produce the model SEDs plotted in Figure~\ref{fig:J1549SED}) are reported in Table~\ref{tab:fits}.
We provide the contour plots of the posterior distributions in the Extended Data Figures~1 and 2.
The reduced $\chi^2$ values are mainly driven by the 1.5 and 141~GHz data points, and our limited degrees of freedom. Our other metrics testing for conversion and goodness of fit provide confidence in our models (See Section~\ref{sec:method_sed} and Supplementary Section~1). 
We directly measure the amplitude of the tSZ effect, giving us the integral of the electron pressure over the volume of the hot gas, $\int \mathrm{p_e dV}$.
Our model results produce a total thermal energy of (9.7$\pm$1.6)$\times$10$^{60}$~ergs according to Equation~\ref{eq:Etherm}.
This measurement of the spectral distortion in the SED indicates that there is localized, hot, tSZ-producing gas at least on scales (10-15~kpc) at the location of J1549B.
We explore an alternative model and its physical implications in Section~\ref{sec:method_sed}, Supplementary Section~3, and show the posterior distributions of that model in Extended Data Figure~3. 


\begin{table*}[!t]
    \centering
    \caption{J1549+1245 System parameters. The colums are the source name and SED fit parameters: synchrotron slope, $\alpha$, total thermal energy of the tSZ producing gas, log of the total infrared luminosity, dust temperature, dust emissivity. Uncertainties are drawn (and propagated, where necessary) from the 16th and 84th percentiles of the marginalized parameter posterior distributions. If no uncertainty is listed, the value is fixed in the model. }
    \label{tab:fits}
        \begin{tabular}{lcccccc} 
		\hline
		Name & $\alpha$ & E$_\mathrm{th}$ & log$_\mathrm{10}$(L$_\mathrm{IR}$/erg~s$^{-1}$) & T$_\mathrm{d}$ (K) &  $\beta$ & $\chi^2_\mathrm{red}$ \\
         & & ($\times$10$^{60}$~ergs) & & & & \\
		\hline
		J1549A & -0.18$_{-0.09}^{+0.09}$ & - & 46.4$_{-1.0}^{+0.3}$ & 56$_{-20}^{+17}$ & 1.7 & 1.03\\
        \\
		J1549B &  -0.61$_{-0.04}^{+0.04}$ & 9.7$_{-1.6}^{+1.6}$ & 46.2$_{-1.2}^{+0.6}$ & 48$_{-20}^{+24}$ & 1.6$_{-0.4}^{+0.4}$ & 4.6 \\
        \\
        J1549B &  -0.61$_{-0.04}^{+0.04}$ & 9.5$_{-1.4}^{+1.4}$ & 46.2$_{-1.3}^{+0.6}$ & 45$_{-18}^{+24}$ & 1.7 & 2.2 \\
		\hline
	\end{tabular}
\end{table*}


\subsection{J1549+1245 System Environment}

\begin{figure}[!htbp]
\centering
    \includegraphics[width=\columnwidth]{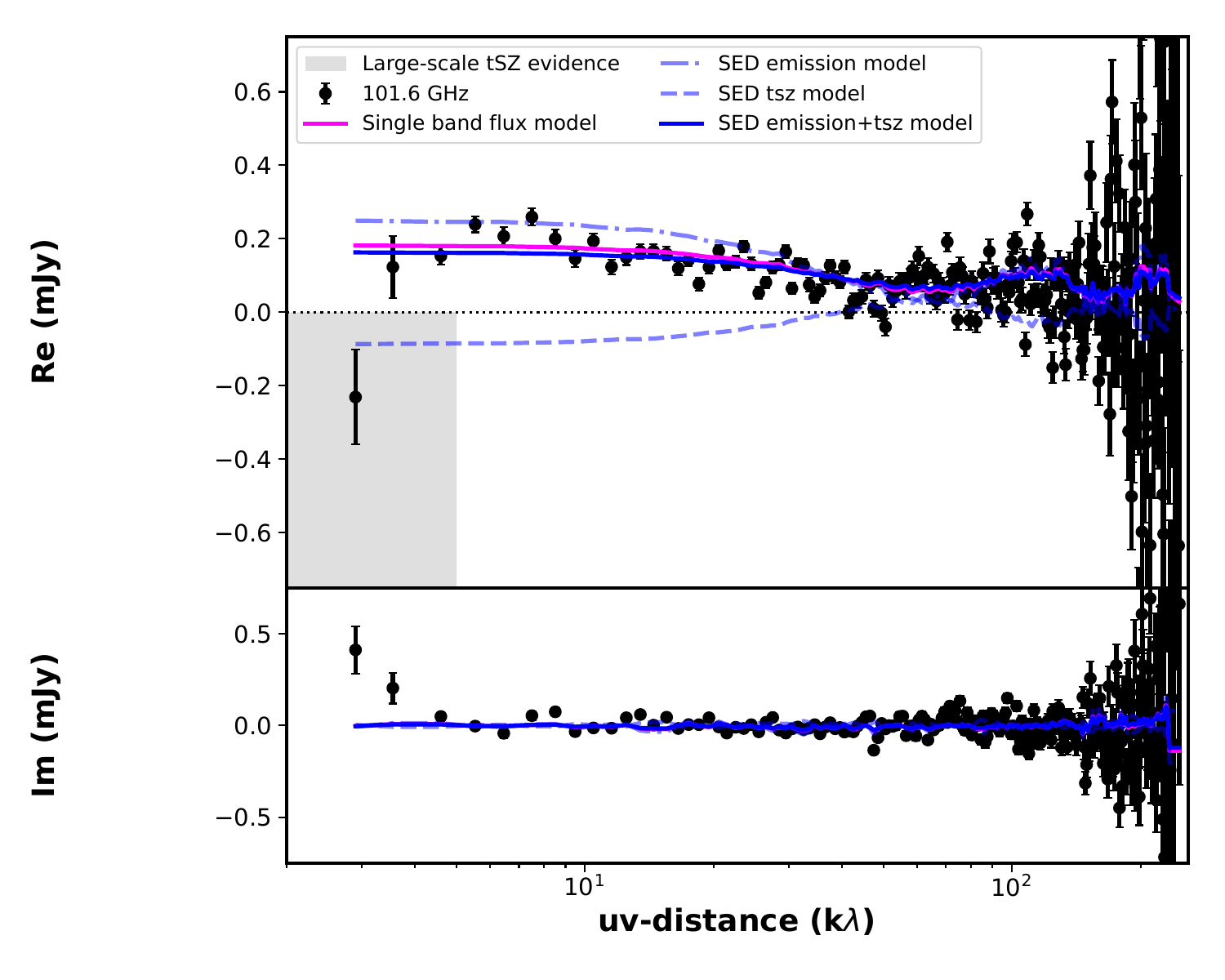}
   \captionof{figure}{Real and imaginary visibilities vs. uv-distance for the J1549+1245 system from the combined 101.6~GHz ALMA and ACA observations. Data are the inverse variance weighted means per bin and error bars represent the square root of the weighted variance. The magenta line is generated from the median values of the parameter posterior distributions in the MCMC analysis to model the point source fluxes from J1549A and J1549B. This is a single-band fit to the visibilities. The solid blue curve is generated using the SED model amplitude of the positive emission as the point source fluxes (blue dot-dashed line) plus the SED model tSZ amplitude (dashed line) and projecting them onto the visibility plane using the uv coordinates. The downturn at small uv-distances cannot be accounted for by source emission, and instead matches the shape of the expected decrement profile from the tSZ effect on large scales.}
    \label{fig:J1549uvfit}
\end{figure}

The SED model provides a measure of the hot gas on scales of at least 10-15~kpc, but with the interferometric observations, we probe the tSZ on circumgalactic scales as well.
We model the uv-plane data for the J1549+1245 system at 101.6~GHz. 
The 101.6~GHz data are the concatenation of the calibrated ALMA 12-m and ALMA Compact Array (ACA) measurement sets. The ACA data provide additional coverage at small uv-distance ($\sim$2~k$\lambda$), meaning very large physical scales ($>$500~kpc). 
Figure~\ref{fig:J1549uvfit} shows the visibilities as a function of the uv-distance of these concatenated data binned by 1k$\lambda$. 
The real and imaginary visibilities are fit by a two point-source emission model with amplitudes and RA, Dec coordinates for J1549A and J1549B as free parameters.
The magenta curve indicates this model fit to the full visibilities (no binning) for the point source fluxes. 
The blue dot-dashed line is generated from using the SED-based fluxes from the only positive emission components of the models, the SED-based galaxy-scale tSZ amplitude as a point source decrement is indicated by the blue dashed line, and the combined SED-based amplitude of the emission plus the tSZ decrement is shown as the blue solid line.
The model amplitudes are applied to the full visibilities and binning is performed at the end in the same way for all scenarios.

\begin{figure}[!htbp]
\centering
    \includegraphics[width=\columnwidth]{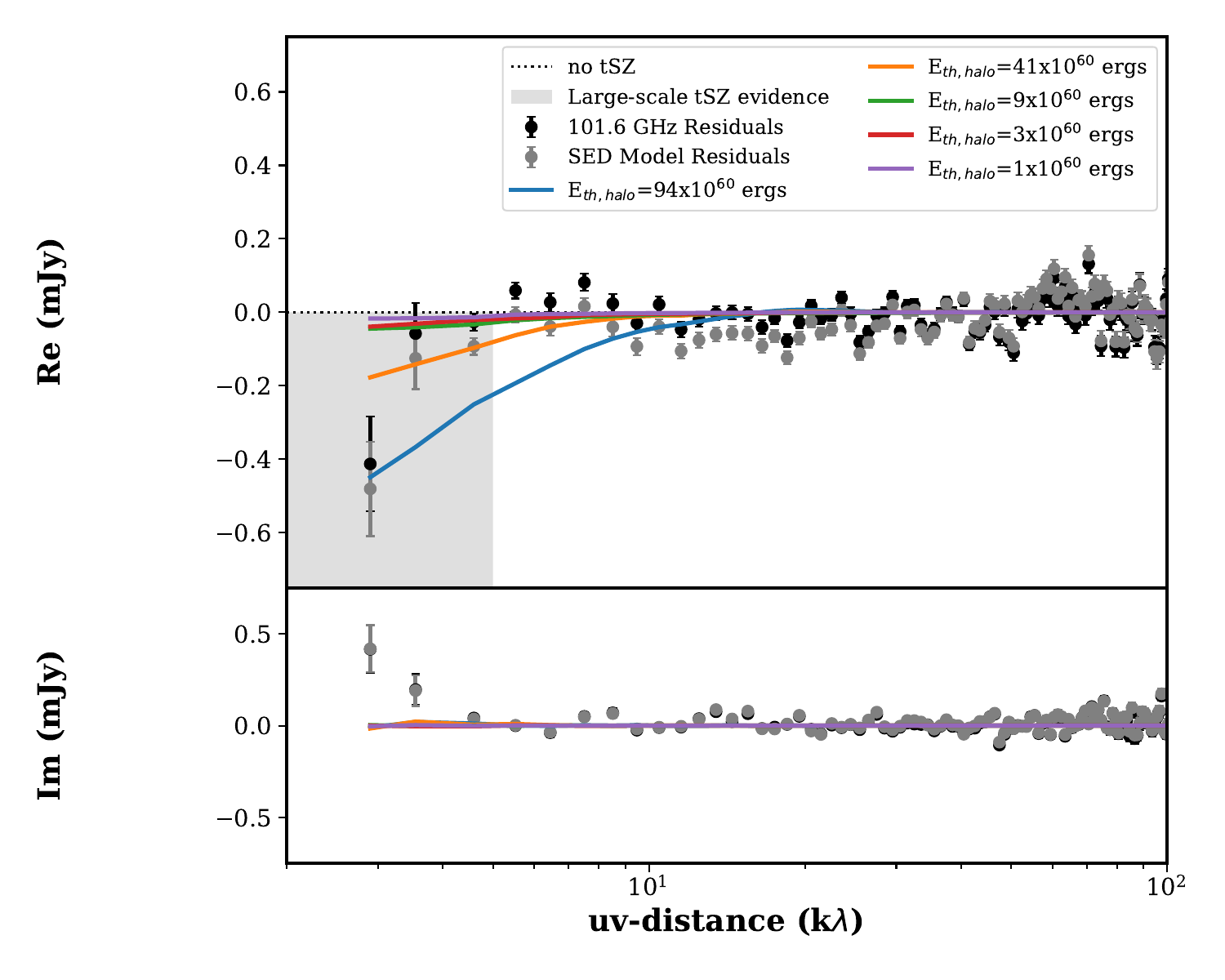}
    \captionof{figure}{Real and imaginary visibilities residuals vs. uv-distance. Black points: visibilities residuals after subtracting the model curve describing only the single-band flux from two point sources (Figure~\ref{fig:J1549uvfit} magenta line). Gray points: visibility residuals after subtracting the SED-based emission model curve (Figure~\ref{fig:J1549uvfit} blue dot-dashed line). The colored lines show the uv-distance profiles of the tSZ from \textsc{fable} halos with the labeled total thermal energies projected onto the same visibility plane as our observations and then binned in the same way as the data. All points are inverse variance weighted means minus the corresponding model and error bars are the square root of the weighted variance.}
    \label{fig:J1549uvfitresid}
\end{figure}

The residuals of these data minus the models are shown in Figure~\ref{fig:J1549uvfitresid}.
At uv-distances less than $\sim$10~k$\lambda$, we begin to observe significant deviations from the two point-source model (the y=0 line of the real visibilities). 
Below 5~k$\lambda$, we observe negative flux densities consistent with what we expect from the tSZ effect on physically large-scales. 
There is a small positive signal at 5-10~k$\lambda$ in the real visibility residuals after single-band source subtraction. 
This may be due to a slightly extended nature of one of the sources (A or B).
Despite this positive excess, we find that a model made up of two point sources produces the best fit to these single-band data (as opposed to a 2D Gaussian shape for one or both sources). 
There is a negative flux measurement of (-0.41$\pm$.14)~mJy in the 2-3~k$\lambda$ bin ($\sim$70 arcsec/590~kpc) of the real component of the visibility residuals. 
The structure of the visibilities is fully consistent with bright central sources accompanied by a larger-scale decrement in flux, apparent in the down-turn of the real visibilities at small uv distances in both the raw data and the model-subtracted data \cite{brow19}.
The significant non-zero amplitudes of the imaginary components at small uv-distances may indicate asymmetries in the structure of the source(s), and may also indicate incomplete sampling of the uv-plane. 
Such incomplete sampling would not change the amplitude of the real component of the visibilities, but can inflate the uncertainties. 
After accounting for the intrinsic scatter in the bin, we determine the significance of the decrement in the smallest uv bin to be 2.9$\sigma$ decrement and caution its over interpretation. 
For comparison, we overlay the uv-distance profiles from \textsc{fable} simulations (\ref{sec:fable}) at the same frequency that we projected onto the same visibility plane as our observations, and then binned in the same way as the data. 
The different lines correspond to the total thermal energy computed from the simulated maps. 
These are not fit to the data, but instead show how the \textsc{fable}-simulated total energy due to SZ in similar haloes compares to our data.

\begin{figure*}[!htbp]
\centering
    \includegraphics[width=\textwidth]{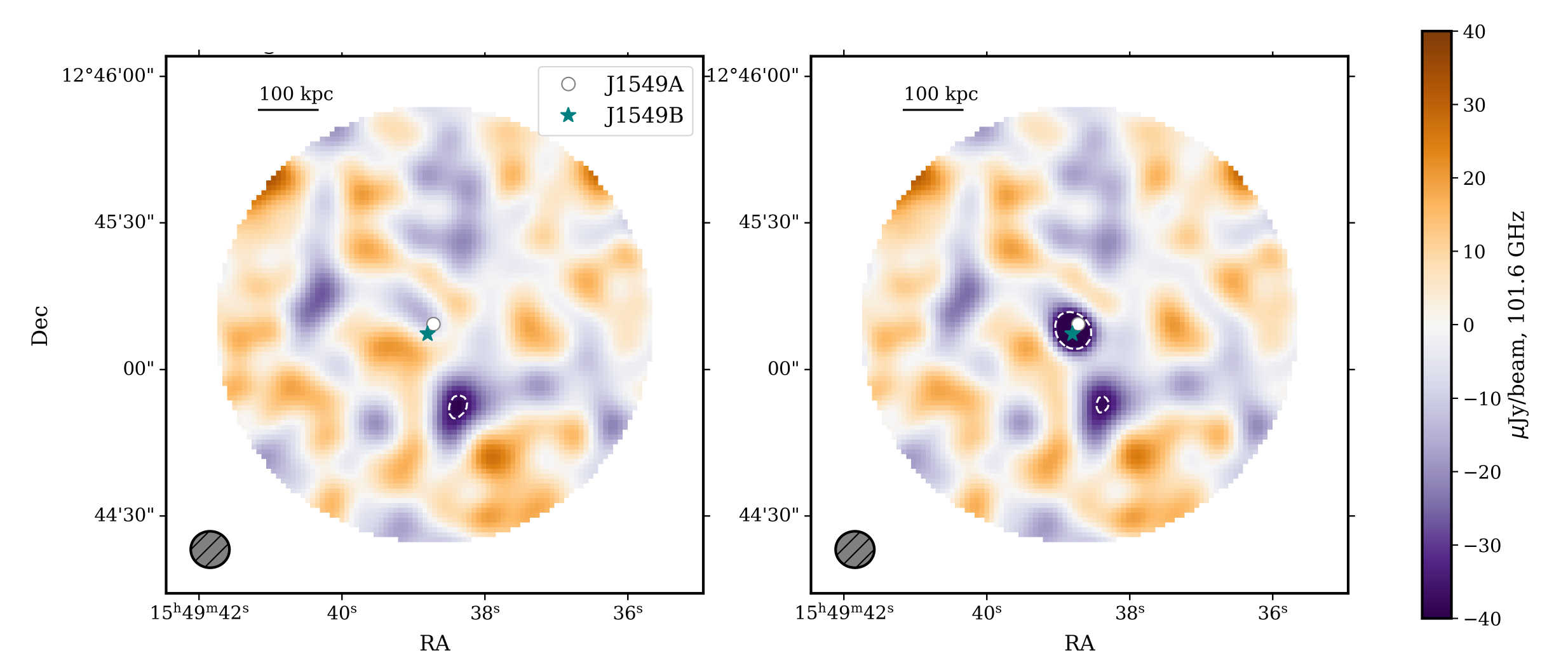}
    \captionof{figure}{J1549+1245 System residual images. The combined ALMA and ACA 101.6~GHz images have an 8~arcsecond uv-taper applied after subtracting the (left) single-band point source flux subtraction and (right) positive emission from the SED model. Contours are -4$\sigma$, where $\sigma$=9$\mu$Jy/beam. The tapered beam is shown on the lower left.}
    \label{fig:J1549_97GHzresid}
\end{figure*}

We generate a measurement set from the 101.6~GHz visibility residuals and image them with a uv-taper of 8~arcseconds to look for a signature of the large-scale decrement in the image plane. 
We observe a region of negative flux with a 4.6$\sigma$ peak flux decrement in the tapered residual map. 
The center of this region is $\approx$100~kpc southwest of the quasar system that is roughly 100~kpc across and cannot be well described by a 2D-Gaussian shape (Figure~\ref{fig:J1549_97GHzresid}, left and right). 
The integrated flux density of the ``bubble" is (-65$\pm$13)$\mu$Jy, a 5$\sigma$ signature of the tSZ in the halo environment.
This ``bubble" is distinct from that probed by the uv-distance plots, which indicate a hot halo of tSZ-producing gas on scales on the order of the primary beam size of the observations (600~kpc). 
This ``bubble" is instead a localized signature of hot gas detectable within the uncertainties of the imaged residuals. 
Figure~\ref{fig:J1549_97GHzresid} shows the residuals after subtracting only the single-band point source fluxes from the visibilities and re-imaging, and the results of subtracting the SED model-based positive emission fluxes from the visibilities to visualize the localized tSZ signature centered on J1549B along with the larger-scale signal. The significance of the 100kpc bubble is unchanged by this subtraction.


\section{Discussion}
\label{sec:discuss}

The physical relationship between the total flux of the tSZ signal and the total thermal energy of the gas is given in Section~\ref{sec:method_sed}, equations~\ref{eq:Ssz}-\ref{eq:Etherm}.
The relationship holds true for the total flux measured on each of the three different scales: the SED model on 10-15 kpc scales over the galaxy, the largest scale negative flux indicated by the uv-distance, and the 100~kpc scale negative bubble indicated in the residual image.
The total thermal energy of the hot gas that produces the tSZ spectral distortion of the SED of J1549B is  (9.7$\pm$1.6)$\times$10$^{60}$~ergs. 
We also detect the negative signal in the residuals of the real visibilities of the system, after subtracting the curves describing the positive emission at 101.6~GHz. 
The amplitude of this tSZ measurement in the smallest uv-distance bin is (-0.41$\pm$0.14)~mJy, which at this frequency corresponds to a total thermal energy of (4.0$\pm$1.3)$\times$10$^{61}$~ergs. 
The negative flux region on 100~kpc scales in the image plane of the residuals produces an integrated tSZ decrement flux density of (-65$\pm$13)~$\mu$Jy, which at this frequency indicates a total thermal energy of (6.4$\pm$1.3)$\times$10$^{60}$~ergs. 
There is another 3$\sigma$ negative flux decrement observed in the residual map, which may be a tSZ signature in the hot halo, but given the noise limits of the reconstructed map, we keep our assessment based on only the more significant flux decrement.
We searched for an intervening cluster or group of galaxies that might produce this signal at 100~kpc, but there does not appear to be any significant visible structure in the Dark Energy Spectroscopic Imaging Legacy Survey \cite{dey19} within the 3$\sigma$ contours of our decrement.

Our goal is to understand the source of the hot gas responsible for the tSZ effect and what implications it has for the role of quasar feedback in this system. 
It is possible that the spectral distortion is caused by the hot gas produced by a combination of all of the scenarios explored herein. 
We have an additional source of information about J1549A via its rest-frame optical spectrum, which reveals outflows of ionized gas evidenced to be significantly blue-shifted to nearly the same systemic velocity of J1549B.
This spectrum was obtained by placing a 1" slit at the location of J1549A oriented with a PA of 0 degrees North, so the [OIII] emission is known to be spatially co-located with the J1549A and has a measured redshift of $z_\mathrm{OIII} = 2.365$ \cite{bisc17}. The line has FWHM=1250~km/s, and the model determining the redshift requires a narrow core component.
The redshift of J1549A as determined by the molecular gas, which is more likely to be bound to the host galaxy, is $z_\mathrm{CO, A}$=2.374 (See Supplementary Section~2, and Supplementary Figures~3 and 4). 
We deduce that the [OIII] centroid in J1549A is up to 800~km/s blueshifted from its molecular gas.

In contrast, the redshift of J1549B determined by the molecular gas is $z_\mathrm{CO, B}=$2.363, only 178~km/s offset from the [OIII] velocity center. 
The ionized gas traces the edge of the quasar wind bubble \cite{gill17}. 
Its presence at this velocity separation from the J1549A-bound molecular gas, provides evidence that the quasar winds extend at least to the distance of J1549B. 
We suggest that the hot gas responsible for the tSZ effect may plausibly arise from either J1549A or J1549B outflows, and we explore the physical interpretation of each scenario.  
In the event that the outflowing wind from J1549A is solely responsible, it would also cause a kinematic SZ (kSZ) effect from the overall bulk flow of the gas \cite{suny72}. 
Because this scenario implies that the outflow is blueshifted, the amplitude of the kSZ would be positive, indicating an even higher total thermal energy (flux decrement) to compensate. 
For the typically high values of quasar outflow velocities and temperatures \cite{fauc12}, the expected contribution to the total SZ effect from the kSZ would be of order 10\% \cite{carl02}.

We first examine the total thermal energy in the context of the bolometric luminosity of J1549A: 5$\times$10$^{47}$~erg/s (\cite{bisc17} and our own computation from with a bolometric correction from \cite{rich06}).
If we assume the decrement around J1549B is a result of the powerful outflows of J1549A and a quasar lifetime of 100~Myr (\cite{cava10}, a conservative upper limit \cite{harr24}), this indicates a coupling fraction $f$ of the quasar winds propagating to the surrounding medium at an efficiency of at least 0.6\% to produce the localized signal probed by the spectral distortion of J1549B's SED. 
Alternatively, we also determined that J1549B hosts a powerful active galactic nucleus (AGN) with a bolometric luminosity of 2$\times$10$^{47}$~erg/s based on the near-IR emission decomposition (Section~\ref{sec:fextract}) and it may itself be responsible for the tSZ spectral distortion seen in its SED. 
Then, for the same conservative upper limit on the quasar lifetime, we find that the total thermal energy of the hot gas seen via tSZ is at least 1.4\% of the bolometric luminosity of J1549B. 
The total thermal energy required to produce the tSZ effect of this amplitude is unlikely to be due to star formation activity.
The total thermal energy of starburst-driven galactic winds ranges roughly from 1$\times$10$^{54}$-1$\times$10$^{58}$~ergs \cite{mcca87,stri00}.
The theoretical maximum energy of the winds is therefore more than two orders of magnitude smaller than that any of our measurements of the total thermal energy of the tSZ-producing hot gas bubbles.

We postulate that the amplitude of large-scale (both the 100~kpc and tentative $\sim$600~kpc) tSZ signals are a result of a combination of hot gas in the massive, assembling halo and from the long-term energy injection from the quasars.
If we consider the entirety of the total thermal energy as probed by the tSZ to be due to the output from the quasars over their lifetime, we can further estimate the upper bound of the coupling fraction of quasar winds to their large-scale environment.
One approach is to consider the cumulative thermal energy probed by the SED signal and the 100~kpc signal,  1.6$\times$10$^{61}$~ergs. 
If the entirety of that signal is due to energy output from both quasars, there is an average coupling fraction of 0.7\%. 
If the entirety of the 600~kpc scale decrement is 
due to the combined bolometric output of the two sources over a 100~Myr time period, we find a coupling fraction of 1.8\%. 

We compare to the \textsc{fable} cosmological simulations \cite{hend18} as a tool to explore the respective contribution from structure formation and feedback (Section~\ref{sec:fable}).
We simulate seven cluster progenitor haloes, with properties similar to the J1549+1245 system, with and without black hole feedback turned on. 
The systems have halo masses ranging from 6.07$\times$10$^{12}$~M$_\odot$ - 5.13$\times$10$^{13}$~M$_\odot$, with the upper end of this mass range estimated to be too massive to describe our system. The halo radii, R$_{200}$, range from 164-334~kpc.
With AGN feedback on, the total thermal energy ranges from 1.24$\times$10$^{60}$-9.4$\times$10$^{61}$~ergs, with an average of 9.6$\times$10$^{60}$~ergs. 
The average total thermal energy inside the entire volume of the hot haloes of systems similar to ours is the same as our detection on the $\sim$12~kpc scales probed by J1549B's SED. 
The simulated system with total thermal energy that best matches our tSZ signal on the largest scales, 4.1$\times$10$^{61}$~ergs is shown by the orange visibility profile in Figure~\ref{fig:J1549uvfitresid}, though there is a discrepancy in the smallest uv-bin. 
This could suggest a difference in the concentration of the SZ-producing gas between the observations and simulations.  
Simulated systems with no quasar feedback, but which still approximately match the measured dynamical masses, reside in dark matter halos with M$_{200} \approx$1$\times$10$^{13}$~M$_\odot$. 
The total thermal energies of these haloes are approximately 2.5$\times$10$^{60}$~ergs, only 26\% of our most significant measurement. 
Similarly, \cite{plan15} report a  linear relation between the integrated tSZ signal and halo mass for halo masses down to 3$\times10^{13}$M$_\odot$; extrapolating it down to 1$\times10^{13}$~M$_\odot$ as in \cite{cric16} produces a total thermal energy of 3.4$\times$10$^{60}$~ergs.
Both of these values are cumulative over the entire halo, and these values are only $\leq$35\% of the tSZ component that we measure around J1549B on 10-15~kpc scales.

Our results represent the first strong ($>$5$\sigma$) detection of the total thermal energy of the hottest phase of gas in an individual quasar system. 
Recently, a similar measurement was made around a z=4.3 known protocluster hosting 3 radio loud quasars \cite{zhou26}. 
We find compelling evidence that the quasar outflows are coupling to the surrounding medium to produce the tSZ with efficiency that is high enough to match that which is required by cosmological simulations to reproduce observed galaxy properties \cite{choi12}.
These results also indicate that this hottest phase of the gas likely carries the most energy. 
The kinetic energy measured in the outflow of the ionized gas from J1549A, for example, is only 0.2\% of its bolometric luminosity. 
While our detection may be a result of the energetic output from the supermassive black hole at the center of J1549B, it is not likely that the other gas phases in this source could reach the level of energetics probed by the tSZ as these metrics rarely reach more than $\sim$1\% \cite{vayn21}. 
This detection of the tSZ effect has far-reaching implications for high resolution measurements in high redshift environments for we can better understanding the hottest gas phases in (proto-)cluster, quasar, and highly star forming environments. We are paving the way for the next generation of interferometric (e.g. ALMA and VLA) and single dish (e.g. AtLAST \cite{dima25}) observations to do this work.

\begin{methods}\label{sec11}

\subsection{Continuum Imaging}

All calibration and data reduction were performed using standard ALMA and VLA pipelines in Common Astronomy Software Applications (\textsc{CASA}).
We created images from all of the interferometric data using the \textsc{CASA} module \textsc{tclean}. 
The ALMA data were imaged with the default imaging pipeline, which used the hogbom deconvolver and Briggs weighting with robust=0.5, and in all cases met or improved upon the requested noise thresholds for each project. We tested multiscale deconvolution with scales at 0, 1, and 3 times the beam size as recommended with no change in the fluxes.
The VLA 1.4~GHz image was taken directly from the Faint Images of the Radio Sky at Twenty-Centimeters Survey cutouts. 
The VLA 3~GHz image was taken from the VLA Sky Survey advanced data products service. We use the high quality single epoch continuum image (tt0), which uses the multiscale deconvolver and Briggs weighting with robust=0.5. 
The higher frequency VLA images (6, 10, and 33~GHz) were imaged using the VLA imaging pipeline with standard parameters, with the exception of the weighting for the 10 and 33~GHz images, which we modified to use natural weighting to maximize sensitivity, which only matter for J1549A. 

Using the continuum images, we extract continuum flux densities from our targets of interest in the ALMA, VLA, and SMA datasets, and identify all significant sources of emission, particularly in ALMA bands 3 and 4 which cover frequencies that are in the regime of a flux decrement due to the tSZ effect.
One method of investigating the presence of the tSZ decrement is to analyze the results of source subtraction in the image.
We find six additional possible sources in the band 4 image at the 4$\sigma$ level. We cannot confirm if these sources are at the same redshift as J1549+1245, but quasars are known tracers of large-scale structure \cite{wang15,hall18}. 

Figure~\ref{fig:J1549im} shows the ALMA band 4 continuum image (color scale), with the highest resolution ALMA band 3 (teal), ALMA band 6 (white dashed), VLA X band (black), and VLA C band (magenta dot-dashed) contours overlain.  
We make an image of the combined band 3 ACA data to check the detection of the J1549+1245 system and look for any additional positively emitting sources in the field, but we only use these data in the visibility plane analysis.

\subsection{Flux extraction}
\label{sec:fextract}

Source fluxes from all interferometric data are extracted using CASA's \textsc{imfit} tool. 
This routine extracts flux densities both for unresolved point sources and (partially) resolved sources that are well approximated by a two-dimensional Gaussian by determining the peak flux density and the integrated flux density, respectively.
All of the source fluxes are determined by drawing a region around each distinct source within the image viewer GUI and running the \textsc{imfit} routine within that region, which locates and fits a point source and two-dimensional Gaussian to the positive emission. 
For the band 3 fluxes in the SED, we use only the observations from 2024 of the J1549+1245 system, in which J1549A and J1549B are resolved from each other. 
J1549A is a point source in all images except 6~GHz, and J1549B is a point source in all images except 6 and 232~GHz. 
Uncertainties on the flux values are the quadrature sum of the error in \textsc{imfit} and either 5-10\% flux uncertainty depending on the telescope/observing band. 
All of the maximum recoverable scales are large enough that we do no expect to be missing any flux.

In the WISE data, J1549A and J1549B are marginally resolved from each other, and we implemented forced photometry to obtain their respective fluxes. For each of the four WISE bands (W1, W2, W3, and W4), the fluxes of the two sources were modeled by fitting two two-dimensional Gaussian components. During this process, the centroids of the Gaussians were fixed to the known source positions taken from the $\sim$1" resolution band 3 ALMA data, and the shapes were constrained to be identical for both sources.
This de-convolution process yields values with reasonable uncertainties for each source in W3 and W4, but J1549B is determined to be significantly fainter than J1549A in W1 and W2 such that we consider the photometry at these wavelengths to be upper limits as indicated by the downward facing triangles in Supplementary Figure~1. 

\subsection{Spectral Energy Distribution Modeling}
\label{sec:method_sed}

We use the observed flux densities and SED models to characterize the physical mechanisms producing the emission components in the near-IR through radio.
We compare the fluxes converted to solar luminosities as a function of rest wavelength (Supplementary Figure~1) to starburst templates from \cite{riek09} with star formation luminosities 10$^{11}$, 10$^{12}$, and 10$^{13}$~L$_\odot$. 
We observe that the millimeter-wavelength portions of the SEDs of both J1549A and J1549B are well matched to the 10$^{12}$~L$_\odot$ template, while the near-IR and radio flux densities of our targets are well above those expected for any of the starburst galaxies. 
We thus conclude that J1549B is a massive (M$_\mathrm{dyn} \approx$ 3.6$\times$10$^{11}$~M$_\odot$ \cite{bisc21}) dusty star-forming galaxy that also hosts a powerful active galactic nucleus with a bolometric luminosity consistent with that of a quasar (L$_\mathrm{bol}$ = 10$^{47.3}$ erg/s).
The dynamical mass of J1549A as M$_\mathrm{dyn} \approx$ (1.2$^{+1.7}_{-0.5}\times$10$^{11}$~M$_\odot$ as traced by CO(4--3) \cite{bisc21}, and we use the entire dynamical mass of the system 4.8$^{+2}_{-0.6}\times$10$^{11}$~M$_\odot$ in our comparison to simulations and in our interpretation of the implications of our measurements, with a wide range of uncertainty in these masses and their associated halo mass. 

In Supplementary Figure~2 and Figure \ref{fig:J1549SED} we plot the flux densities versus observed frequencies for J1549A and J1549B from the VLA, ALMA, and SMA (upper limit, J1549B only).
We over-plot each of their respective models in magenta and cyan.
The models are generated using the affine-invariant Markov Chain Monte Carlo (MCMC) \texttt{emcee} sampler \cite{fore13}.
We model the radio component as a power law with respect to frequency, $S_{\nu} \propto \nu^{\alpha}$. We impose a uniform prior on $\alpha$ bounded by  $-1.5 < \alpha < +0.1$, consistent with optically thin synchrotron emission. The normalization amplitude is free to vary above zero.
We allow the spectrum to be flat with a slightly positive $\alpha$ \cite{dezo10}, particularly for J1549A. 
The relatively low flux densities and flat slope between 10 and 32.3~GHz in this source suggest that they are the result of free-free radiation rather than synchrotron, and this is supported by the maximum likelihood model (see Supplementary Section~1).

We are sampling the Rayleigh-Jeans portion of the cold dust emission spectrum and modeling it as
\begin{equation}
    S_{\mathrm{dust}}(\nu, z, L_{\rm IR}, \beta, T_d) = \frac{L_{\rm IR}}{4\pi (1+z)d_M^2(z)} \times \\ \frac{((1+z)\nu)^{\beta}B_{(1+z)\nu}(T_d)}{\int \nu'^{\beta}B_{\nu'}(T_d)d\nu'},
    \label{eq:Sdust}
\end{equation}
where $B(T_d)$ is the Planck blackbody function. 
The temperature of this optically thin modified blackbody spectrum is $T_d$ and its emissivity index is $\beta$. 
The integral is taken from $8-1000$~$\mu$m in the rest frame in order to normalize the spectrum to the infrared luminosity $L_{\rm IR}$ of the quasar over the same wavelength range as is standard practice \cite{kenn98}. 
The comoving distance $d_M(z)$ at the redshift of the quasar is used to generate the SED in terms of the observed frame flux density. 

The tSZ effect produces a distortion of the spectrum of CMB radiation passing through any hot plasma, such as those produced by quasar winds. 
The tSZ effect is proportional to the volume-integrated thermal pressure of the plasma's electron gas, $\int p_e dV$. 
This integrated thermal pressure is directly proportional to the thermal energy of intervening gas, $E_{\rm th}$, in this case the hot plasma produced by the quasar winds. 
The tSZ contribution to the spectral flux density integrated over the solid angle of the source is: 
\begin{equation}
\label{eq:Ssz}
    S_{\rm SZ}(\nu,z, \int p_edV) = I_0 g(x) \frac{\sigma_T}{m_ec^2}\frac{\int p_edV}{D_A^2(z)},
\end{equation}
where $I_0 = 2(k_BT_{\rm CMB})^3/(hc)^2$ and $g(x)$ is the nonrelativistic frequency dependence of the tSZ effect,
\begin{equation}
     g(x \equiv h\nu/k_BT_{\rm CMB}) = \frac{x^4e^x}{(e^x-1)^2}\left( x \frac{e^x+1}{e^x-1} - 4\right).
\end{equation}
The constants $m_e$, $\sigma_T$, and $c$ are the electron mass, the Thomson scattering cross section, and the speed of light. 
$D_A(z)$ denotes the angular diameter distance at the source redshift. 
The total thermal energy of the hot gas bubble then relates to the integrated electron pressure over its volume via
\begin{equation}
    E_{\rm th} = \frac{3}{2}\left(\frac{\mu_e}{\mu}\right)\int p_e dV = fL_{\rm bol}\tau,
    \label{eq:Etherm}
\end{equation}
where we assume solar abundances and a fully ionized gas such that the total mean weight per particle is $\mu=0.613$ and the mean weight per free election $\mu_e = 1.17$.
This formalism projects a localized, finite 3D bubble of gas onto the 2D sky. It allows for the hot plasma from the quasar wind to be contained within a localized, roughly spherical boundary (out to the shock front or the radius of the wind bubble, for example), but the measured flux could be picking up signal along a cylinder through the halo.
We further relate the total thermal energy to the bolometric luminosity output of the quasar over some timescale $\tau$ via the coupling fraction $f$ of that output to produce the total thermal energy.

The amplitude of the tSZ is parameterized as $\int p_e dV$, and is free to vary.
The modified black body parameters are all free to vary with uniform priors for the fit to J1549B's SED. 
The emissivity index $\beta$ is bounded by [1,2.5], the dust temperature is bounded by [10,80]~K, and the $log10(L_{\rm IR}/ergs)$ is bounded by [44.5,48]. These ranges cover that which is found for dusty star forming galaxies and quasars \cite{kirk15}.
We marginalize over the parameters in our model and report the 50th percentiles of the posterior distributions. The reported uncertainties are statistical uncertainties given by the 16th and 84th percentiles of the marginalized parameter distributions.
The modified blackbody parameters are degenerate with each another, affecting the posterior distributions (Extended Data Figures~1 and 2 for the $\beta$ free and $\beta$ fixed runs, respectively). If we fix any one of them to specific values, we find no significant alteration of our results. 
We fix $\beta$=1.7 in our fiducial model of J1549A due to the limitation of free parameters for this source, as explained further in Supplementary Section~1. 
When we fix $\beta$=1.7 in our model for J1549B, the parameter estimates stay the same, with marginal ($<$1$\sigma$) increase significance of the tSZ amplitude after marginalizing over the other parameters.
We test different values of fixed $beta$ across the prior range and find the amplitude of the tSZ changes by $<$1$\sigma$ and that significance of the amplitude ranges from 6$\sigma$ to 6.7$\sigma$ .
We report the results of fitting J1549B SED with alternative model SEDs in the Supplementary Section~3 (See also Extended Data Figure~3). 
In all modeling with \textsc{emcee} we check for convergence of the chains through the plateau of the autocorrelation time for each of the parameters. 

\subsection{Visibility modeling and source subtraction}

There exist several tools that are publicly available for fitting source fluxes and shapes to interferometric visibility data, and we explore the usefulness of many of these. 
The simplest is CASA's \textsc{uvmodelfit}, which only allows the user to fit one source at a time, with the only user choice to either vary the parameters from their initial guess or not. There is no flexibility in the prior.
We use this routine as a check our flux density extraction measurements, finding no change compared to the \textsc{imfit} results, but we do not find this to be an effective tool for subtracting all the sources in our observed field.

We make use of the publicly available python package \textsc{galario} \cite{tazz18} to fit the uv-plane data for the positively emitting sources in the concatenated band 3 ALMA 12m and ACA measurement set. 
This routine requires a table of visibility data, which can be obtained using the CASA `tb' object. 
We use the publicly available \textsc{uvplot}  module \cite{tazz21} function \textsc{export\_uvtable} to generate a table containing the u and v coordinates, the real and imaginary parts of the complex visibilities and their weights.
We input the source coordinates of J1549A and J1549B obtained from \textsc{imfit} as priors to the model in the uv-plane. 
We find that these sources are point-like in the data, so a simple model consisting of amplitude and location as free parameters is sufficient. 
\textsc{galario} uses \textsc{emcee} to perform a maximum likelihood estimation of the source parameters to generate model visibilities on the same uv-plane as the observations, and compare that model to the entire set of visibilities (real and imaginary) without binning. 
\textsc{galario} accepts the visibility weights as uncertainties. The weights are generated as a result of the NRAO pipelines for science ready data products. We display the binned observed and model visibilities, which are inverse variance weighted means over bin sizes of 1~k$\lambda$ with uncertainties given by the inverse square root of the sum of the weights, $\sigma=1/\sqrt{\Sigma w}$. 

After subtracting the model from the data, we determine the negative flux amplitude in the smallest uv-bin. We report the uncertainty on the flux decrement that includes the intrinsic scatter in the visibilities within the uv-distance bin. 
From the model and residual visibilities, we produce residual (source-subtracted) images by generating residual measurement sets. These are copies of the real measurement sets, but we replace the visibilities with those from the residual visibilities, then we can image them with CASA's \textsc{tclean} function.
In the images, we see a significant region of flux decrement.
We investigated whether there is significant correlation between the real and imaginary visibility amplitudes in the 0-3~k$\lambda$ bin as well as between 0-10~k$\lambda$ using a Pearson correlation test, and we find no strong statistical significance ($\rho<$0.01 an p$>$0.5 in both cases). 
We generated phase shifted uv-distance plots using the center of this decrement signal to determine the offset, and we found that the amplitude of the imaginary component in the smallest uv-bin decreases to be consistent with zero, while the real component of the visibilities remains the same. 
This analysis provides further evidence of the significance of the large-scale decrement, and supports the hypothesis that an asymmetry in the sky signal is responsible for the amplitude of the imaginary component. 

\subsection{Image plane analysis}
We search for the tSZ effect directly in the continuum images in the regime of the decrement. 
The quasar winds may create localized bubbles of hot gas that will produce the decrement in the form of negative ``bowls" in the images, or we might expect to measure the long term heating of the surrounding medium in the form of overall average negative flux across the image after proper subtraction of the positive emission from the sources in the field. 
The latter could yield a negative integrated signal or could also appear as localized ``bowls" measurable as distinct from the expected noise in the maps \cite{benn24, brow19, rowe11}.

Such measurements require careful source subtraction in case of contamination from source emission and for clear determination of the noise in the map.
We may expect the hot gas bubbles to produce the tSZ decrement physically coincident with the host galaxy emission, but detecting that co-located signal in this way is not reliable as it may be confused with over-subtraction of the source emission. 
We may expect, however, a decrement just beyond the extent of the boundaries of the galactic emission. 
In this case, there could be either a ring of negative flux surrounding the host galaxy if the tSZ-producing gas is symmetric, or we might expect a decrement signal offset from the host galaxy emission, as was seen in \cite{lacy19, brow19}. 

We observe the latter in our ALMA+ACA image, even before source subtraction, at the $>$4$\sigma$ level. We generate a reconstructed image after source removal using the point-source subtracted visibilities.
We apply an 8 arcsecond uv-taper to the image to down-weight the visibilities smaller than this scale and emphasize any large-scale signals in the map.
In order to measure the integrated flux of our measured decrement in the reconstructed image, we include all emission within a 2$\sigma$ contour around the bowl in the primary beam corrected map, and compute the uncertainty on the integrated flux as $\sqrt{N}\times$RMS, where $N$ is the number of beams covered by the integrated area.  
The integrated flux value reported in the main text is from the single-band point source flux subtraction. Subtracting the visibilities from the positive emission from the SED model and re-imaging produces differences in the residual image, and the large-scale decrement has 4.8$\sigma$ significance in that version. We note that this is based on two different modeling schemes and thus find the value based on the fit to the uv-plane emission amplitude a more reliable value. 

We further test the statistical significance of this bubble by quantifying the likelihood of randomly generating a decrement at this 5$\sigma$ integrated flux level. 
That is, we quantify the probability of generating a feature with the observed integrated flux from random, correlated noise.
We do this by randomizing the phases of the measurement set; we multiply each visibility by a random phase. This preserves uv coverage, amplitude distribution, and thermal weight structure, but removes coherent source signal.
Over 600 bootstrap realizations, none of the integrated fluxes over the same area produce a signal of the same strength. Such a test puts the p-value at $<$0.002. 
We also fully reproduce the signal using only half-period visibilities. 

\subsection{Visibility plane analysis}

We perform our visibility plane analysis on the band 3 data, as it is this frequency for which we have both ALMA 12-m and ACA observations that cover small uv-distances and larger physical scales. 
We analyze the concatenated data.
The signature of the tSZ effect will similarly be in the real component of the visibilities in the form of a decrement in flux after the positive emission has been carefully subtracted. A large-scale tSZ signal causes a decrement in flux at the shortest uv-baselines. 
After fitting all the positively emitting sources in the uv-plane, we subtract the model uv-distances from the data to produce residuals (Figure~\ref{fig:J1549uvfitresid}). 
We plot the real and imaginary visibilities vs. uv-distance centered on the phase center of the observations. 
We find our results are not altered by shifting the phase center of the plots to that of the center of either J1549A or J1549B. We do not know the phase center of the tentative large-scale tSZ signal probed by this analysis, so the observation phase center is a reasonable choice.
After subtraction of the positive signal, we find a negative flux in the smallest uv-bin, corresponding to $\sim$600~kpc scales, of (-0.41$\pm$0.14)~mJy. The uncertainty captures the inherent scatter in visibilities binned by 1~k$\lambda$.

We attempted to fit a negative component directly as a Gaussian with negative amplitude in the visibilities, but there are two complications that make this approach not feasible. 
Firstly, we do not know {\it a priori} the scales on which the tSZ is present. 
If the signal is only on scales that are approximately the size of the galaxy, then there is a complete degeneracy between the positive and negative emission between the two flat lines in uv-space that describe point sources. 
This is one reason why the multi-frequency SED modeling is a worthwhile endeavor, as we can model the physical spectrum of the positive emission and determine the decrement in the expected flux from the physics driving the emission mechanisms. 
If, as may be expected, the tSZ-producing gas is present on scales larger than the host galaxy, we do not necessarily expect it to be symmetric or a smooth profile, since the ICM is likely not virialized. 
So approximating the tSZ as a Gaussian with negative amplitude centered on the phase center of the observations is not necessarily expected to yield physical results.

\subsection{FABLE Simulations}
\label{sec:fable}
In this paper we use results of cosmological simulations based on the \textsc{fable} model \cite{hend18}, ran with the \textsc{Arepo} code \cite{spri10}. We run zoom simulations of the halo `c448', selected from the parent box Millennium XXL \cite{angu12}, formerly studied in \cite{benn20}. These simulations have a dark matter mass resolution of $5.54\times10^7\,h^{-1}\,\mathrm{M}_\odot$, and an average baryonic mass resolution of $1.11\times10^7\,h^{-1}\,\mathrm{M}_\odot$. All simulations assume cosmology consistent with \textit{Planck} \cite{plan15}. The fiducial \textsc{fable} simulations use a modified version of the feedback model of Illustris, described in \cite{voge13}. For a full description of the \textsc{fable} model see \cite{hend18}, but we briefly highlight the black hole related parts of the model below. 

In \textsc{fable}, black hole seeds of $10^5\, h^{-1} \, \mathrm{M}_\odot$ are placed at the gravitational potential minimum of haloes above a halo mass threshold of $5\times10^{10}\,h^{-1}$\,M$_\odot$. Black holes can grow through mergers with each other or through accreting gas via a Bondi-Hoyle-Lyttleton-based prescription. The luminosity of an AGN (and therefore the energy potentially available for feedback) is related to the black hole accretion rate via $L = \epsilon_\mathrm{r} \dot{M} c^2$, where $\epsilon_\mathrm{r} = 0.1$ is the radiative efficiency of the black hole used in \textsc{fable}. Due to the radiation output from the black hole, the accretion rate is assumed to be capped at the Eddington limit.

\textsc{fable} uses a dual-mode feedback model for AGN, depending on the black hole accretion rate. For the quasar mode, at high accretion rates, 10 per cent of the feedback energy is isotropically injected as thermal energy to neighbouring cells of the black hole. Additionally, a duty cycle is imposed - feedback energy is stored for 25\,Myr before being injected at once, which can make quasar-mode feedback more effective. At low accretion rates a radio mode is invoked, which injects hot bubbles into the gas around the black hole \cite[following][]{sija07} every time the black hole mass changes by 1 per cent. 

In this work, we look at 7 haloes at redshift $z=2.4$ which are similar in mass within 10~kpc to the summed dynamical masses of J1549A and B. These simulated haloes are the largest progenitors of the halo `c448' at $z=2.4$. These all become part of the central cluster by $z=0$ (which has a mass of $10^{15}\,\mathrm{M}_\odot$), but at $z=2.4$ these all lie at least 1~pMpc from each other and have separate hot haloes. We additionally run the simulation again without any AGN feedback injection to explore how important \textsc{fable}'s AGN feedback is in setting the thermal state of the growing intracluster medium. In the haloes we look at, we find an average 7 per cent enhancement of the total thermal energy of circumgalactic gas within $R_{200}$ at $z=2.4$ with AGN feedback switched on. 

To compare the detected SZ signal to our simulations, we make maps of the SZ decrement (including both thermal and kinetic components, though the thermal signal dominates throughout the halo) from our simulation output projected along the $z$ direction of the box, as previously done in \cite{benn24}, using the equations of \cite{carl02}. 
We then convert the image to visibilities using \textsc{galario}'s \textsc{SampleImage} function and the u and v points from the combined ALMA plus ACA observations at 101.6~GHz. We then bin these visibilities in the same way as the observations for the comparison in Figure~\ref{fig:J1549uvfitresid}.

\end{methods}

\begin{addendum}
\addsec{Data Availability}
All data used in this publication are publicly available via National Radio Astronomy Observatory (NRAO) archives and the Submillimeter Array Radio Telescope Data Center.
The ALMA and VLA data can be accessed via the NRAO's science ready data products portal: https://data.nrao.edu/.
The following are the ALMA project IDs used herein: 2018.1.01806.S, 2019.1.01070.S, 2024.1.00933.S.
The VLA project IDs are the following: 19A-160, 20A-323, 20A-363, 22A-253, 23A-076.
This paper makes use of the Submillimeter Array Project 2020A-S051, which can be found through the public data archive at URL: https://lweb.cfa.harvard.edu/sma-archive/
This research has made use of the VLASS primary beam corrected images VLASS-QL\_1.2\_J154938+124500, VLASS-QL\_2.2\_J154938+124500, and VLASS-QL\_3.2\_J154938+124500, which can be obtained at URL: https://cutouts.cirada.ca, operated by the Canadian Initiative for Radio Astronomy Data Analysis (CIRADA). 
This research has made use of the FIRST (Faint Images of the Radio Sky at Twenty-cm) survey cutout image centered on J1549+1245, which can be found by entering the RA and Dec here: https://sundog.stsci.edu/cgi-bin/firstcutout.

\addsec{Acknowledgements}

The Submillimeter Array is a joint project between the Smithsonian Astrophysical Observatory and the Academia Sinica Institute of Astronomy and Astrophysics and is funded by the Smithsonian Institution and the Academia Sinica. 
The authors wish to express immense gratitude for the use of the Mauna Kea observing site, which is one of the most revered places on the islands of Hawai'i.

The authors acknowledge the use of data from the National Radio Astronomy Observatory which is a facility of the U.S. National Science Foundation operated under cooperative agreement by Associated Universities, Inc. 
The Atacama Large Millimeter/sub-millimeter Array (ALMA) is a partnership of ESO (representing its member states), NSF (USA) and NINS (Japan), together with NRC (Canada), NSTC and ASIAA (Taiwan), and KASI (Republic of Korea), in cooperation with the Republic of Chile. The Joint ALMA Observatory is operated by ESO, AUI/NRAO and NAOJ.
The Canadian Initiative for Radio Astronomy Data Analysis (CIRADA). 
CIRADA is funded by a grant from the Canada Foundation for Innovation 2017 Innovation Fund (Project 35999), as well as by the Provinces of Ontario, British Columbia, Alberta, Manitoba and Quebec, in collaboration with the National Research Council of Canada, the US National Radio Astronomy Observatory and Australia’s Commonwealth Scientific and Industrial Research Organisation.

KRH is grateful for helpful conversations with Tobias Marriage, Luca Di Mascolo, and Tony Mrockowski regarding proposals, high-resolution tSZ imaging, and visibility modeling. 

The authors extend our gratitude to the three external reviewers of this manuscript for their time and expert advice. Their comments improved the rigor and clarity of this work.

\addsec{Author Contributions Statement}
KRH was responsible for project planning, analysis, writing, and editing. NLZ was responsible for project planning, writing and editing. JSB was responsible for data analysis of the simulations, writing, and editing. Y-CC was responsible for data analysis and writing. XF was responsible for project direction and editing.

\addsec{Funding Statement}
KRH acknowledges the Smithsonian Institution for funding her as a Submillimeter Array Fellow while carrying out this work. 
In the early stages of this project, KRH was supported in part by the National Science Foundation Student Observing Support award SOSPA6-013. 
JSB acknowledges support from the Simons Collaboration on Learning the Universe and a Leverhulme Trust Early Career Fellowship.

\addsec{Statement of Competing Interests}

The authors declare no competing interests.

\end{addendum}

\onecolumn                
\extendeddata

\begin{figure}[!htbp]
\centering
        \includegraphics[width=0.75\columnwidth]{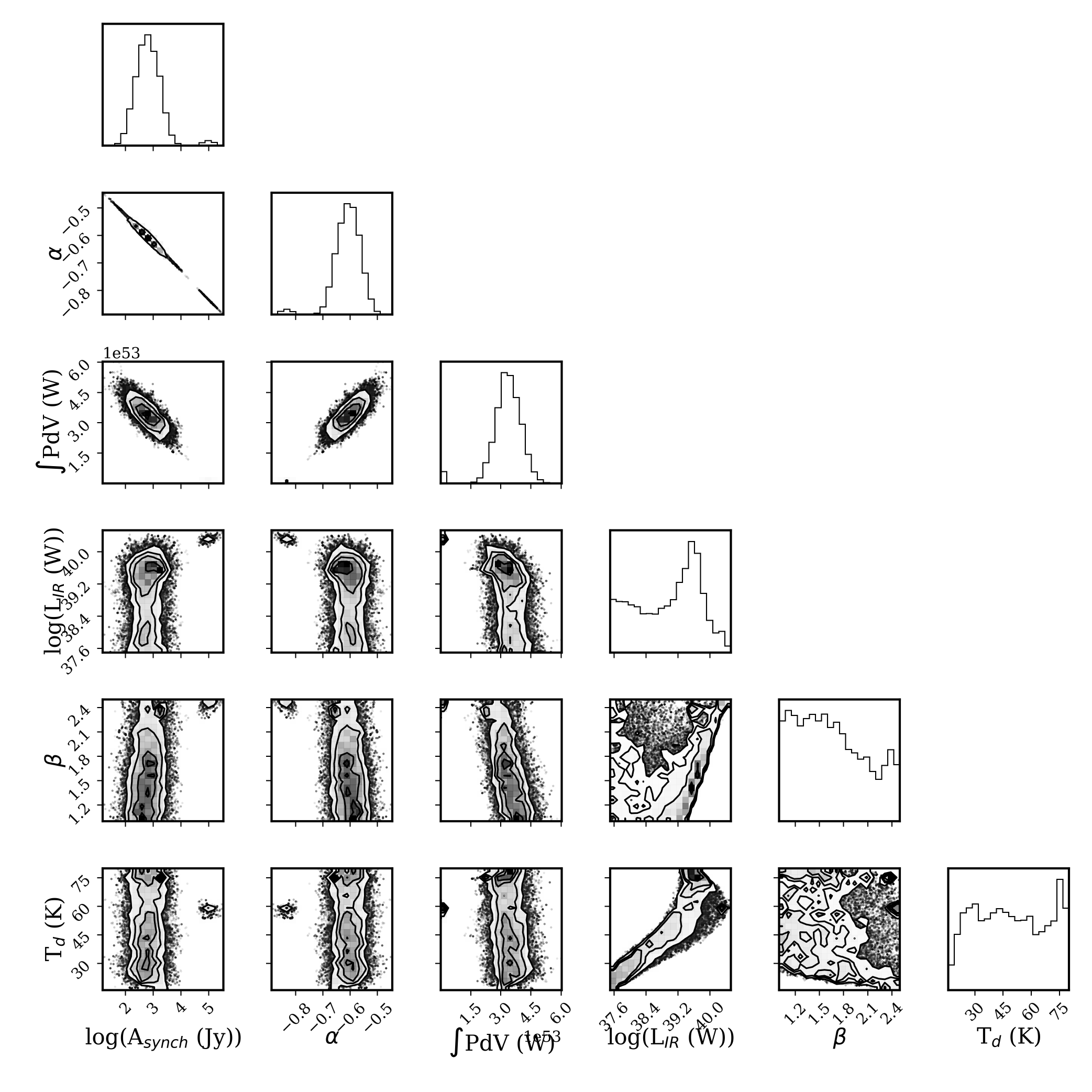}
    \caption{Contours of the parameter samples for the fiducial model fit to J1549B SED. In this model, $\beta$ is a free parameter. The histograms display each parameter’s marginalized posterior distribution. It is not surprising that the dust parameters are not well constrained as there are strong degeneracies among them, and we are only sampling the Rayleigh-Jeans tail with our data points.}
    \label{fig:J1549B_corner}
\end{figure}

\begin{figure}[!tbp]
\centering
        \includegraphics[width=0.75\columnwidth]{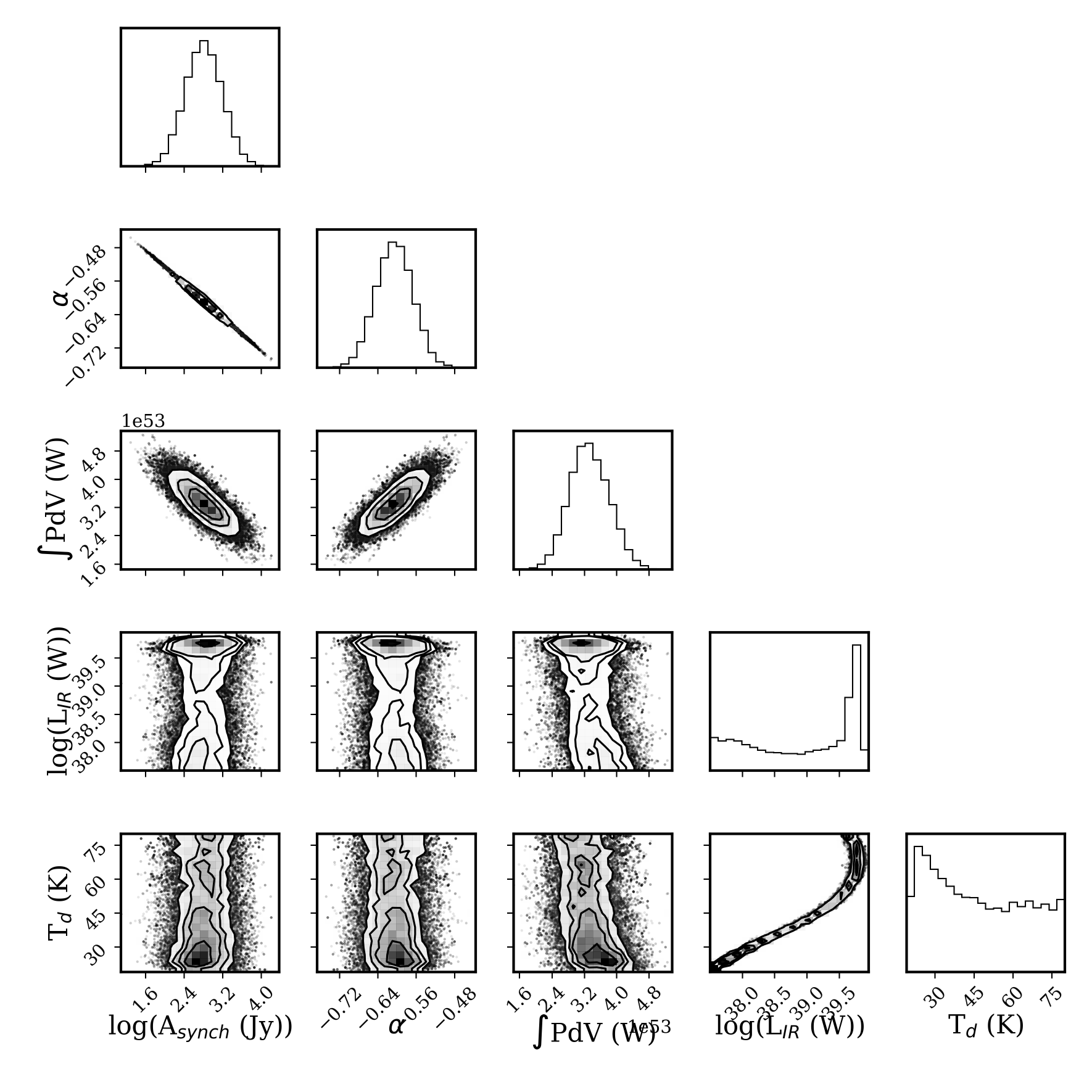}
    \caption{Contours of the parameter samples for the secondary model fit to J1549B SED. The parameter $\beta$ is fixed to 1.7 in this model. The histograms display each parameter’s marginalized posterior distribution.}
    \label{fig:J1549B_corner_betafix}
\end{figure}

\begin{figure}[!tbp]
\centering
        \includegraphics[width=0.75\columnwidth]{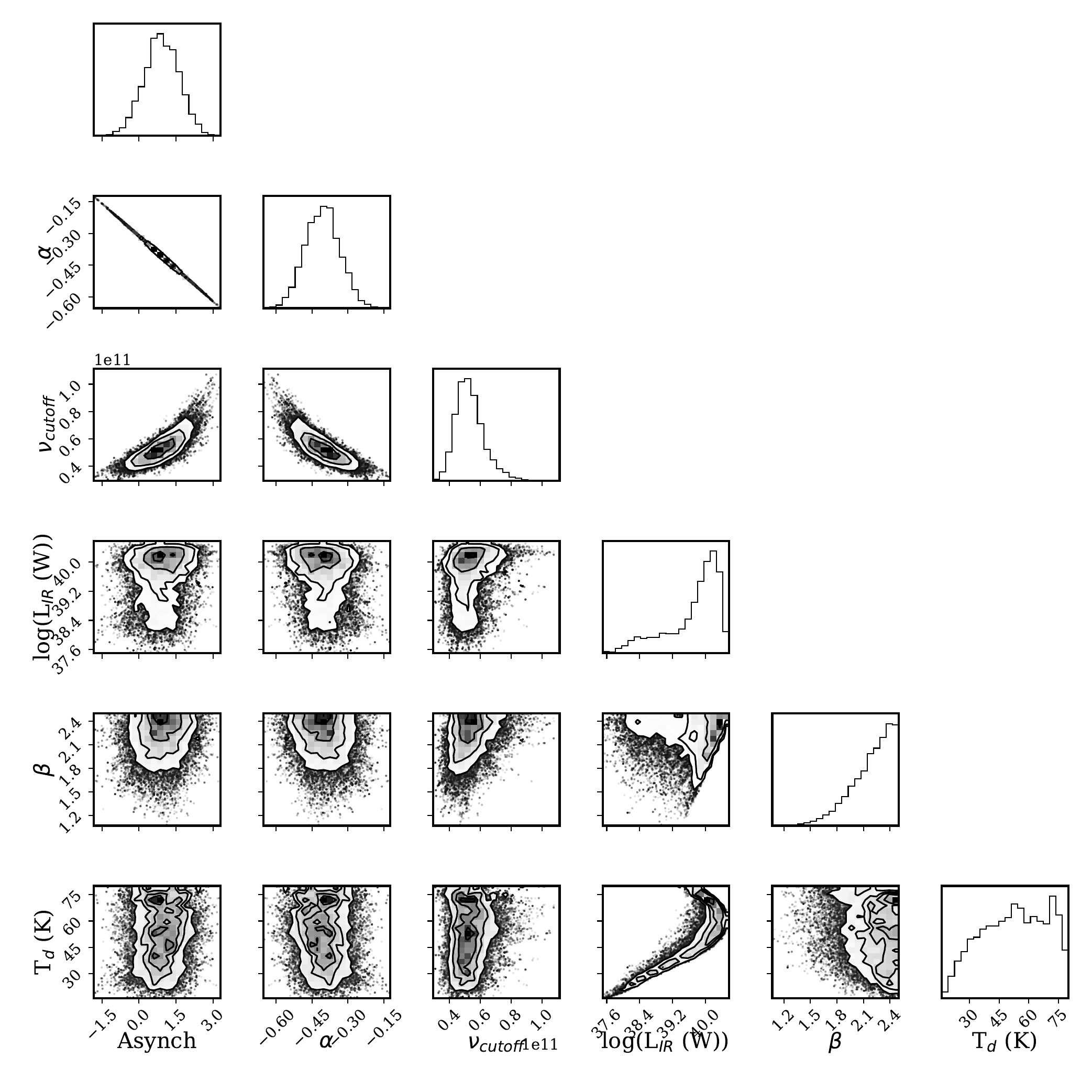}
    \caption{Contours of the parameters samples for the exponential cutoff model fit to J1549B SED. This model uses an exponential cutoff in the synchrotron spectrum and no tSZ. The histograms display each parameter’s marginalized posterior distribution.}
    \label{fig:J1549B_corner_expPL}
\end{figure}

\supplementary

\section{Supplementary Information}

\tableofcontents 

\newpage


\section{J1549+1245 Dual AGN System}
\label{sec:J1549sys_discuss}

\begin{figure}[!hbp]
\centering
        \includegraphics[width=\columnwidth]{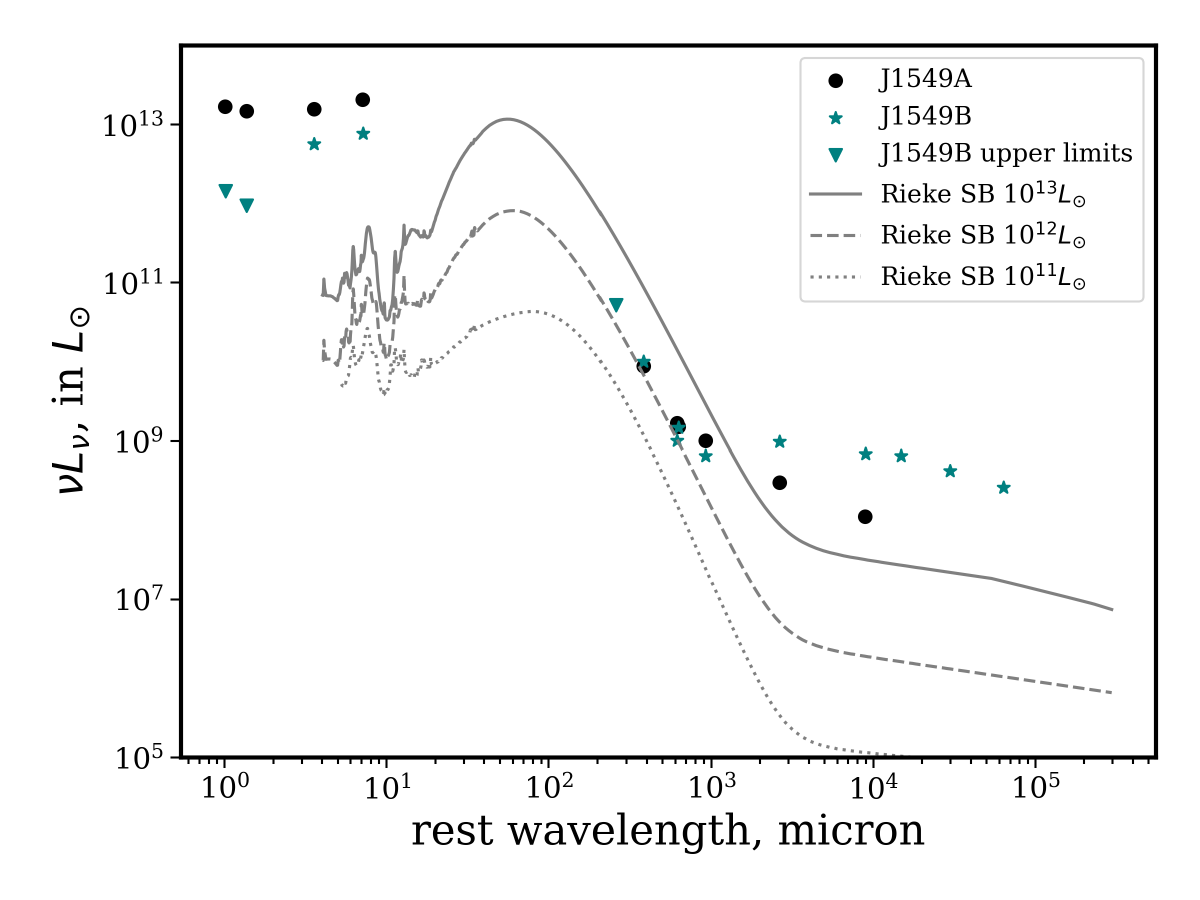}
    \caption{Spectral energy distribution for each galaxy in the J1549+1245 system, expressed in solar luminosities. Black dots are for J1549A and teal stars and triangles are for J1549B. J1549B upper limits are marked as downward pointing triangles. We overlay starburst galaxy templates with three different bolometric infrared luminosities from \cite{riek09} to demonstrate that the radio and near-IR luminosities of neither galaxy are well fit by a purely star forming SED.}
    \label{fig:J1549SEDnuLnu}
\end{figure}

\begin{figure}[!tbp]
\centering
        \includegraphics[width=\columnwidth]{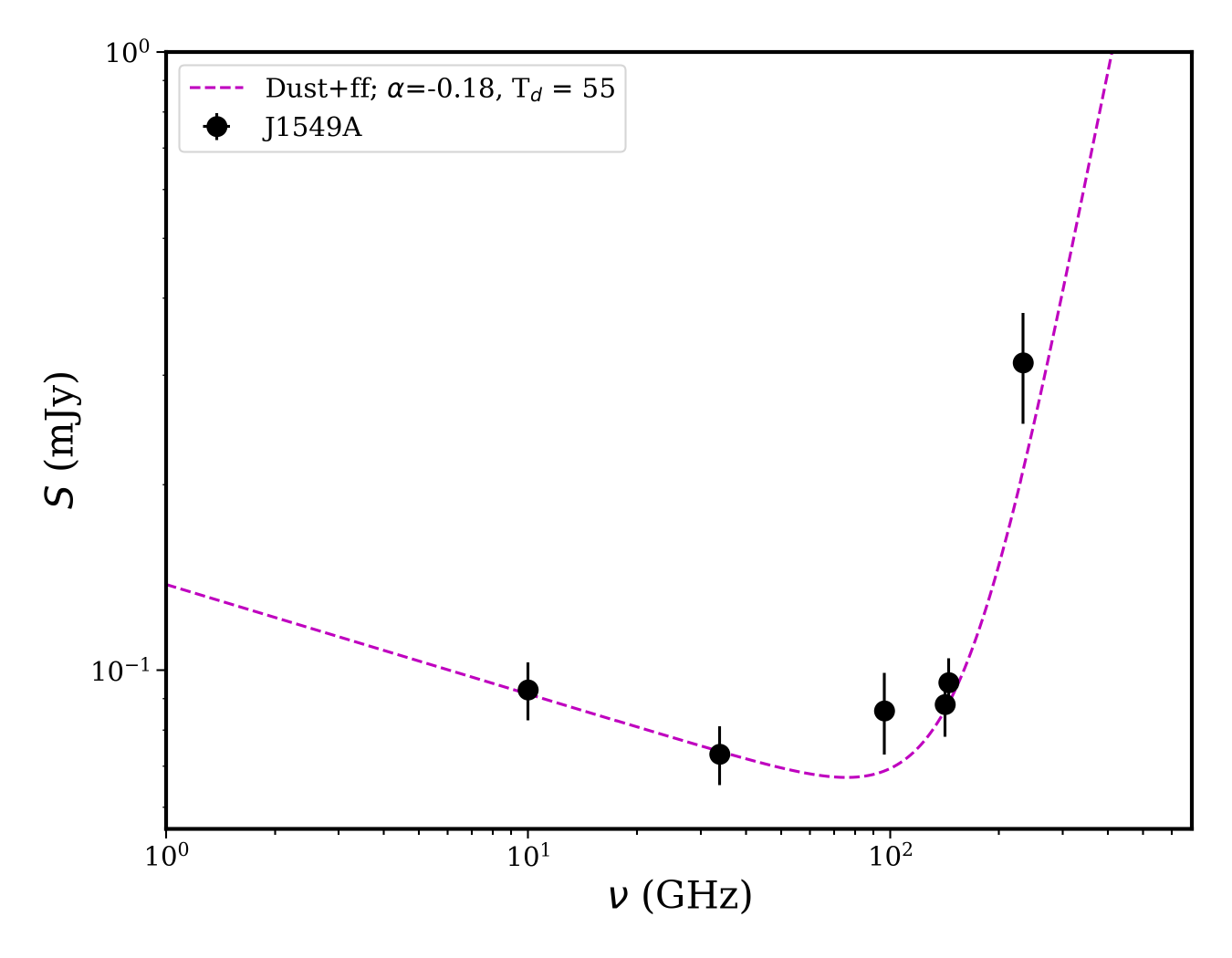}
    \caption{SED for J1549A. The dashed magenta line is our model generated from the medians of the posterior parameter distributions describing the free-free plus cold dust, modified blackbody spectrum.}
    \label{fig:J1549ASED}
\end{figure}

We made use of existing and targeted observations in the radio through sub-mm of the J1549+1245 system in order to understand the physical mechanisms producing the emission. 
J1549A is detected in five bands between 10 and 233~GHz, and the combined source fluxes are measured in the observed near-IR with WISE, which we further decompose (Figure~\ref{fig:J1549SEDnuLnu}). 
We model the radio-sub-mm SED of J1549A with a shallow power-law describing the radio emission, which is likely free-free radiation, plus a modified blackbody to describe a cold dust emission spectrum with an emissivity index $\beta$ fixed to a value of 1.7. The typical spectral index for thermal Bremsstrahlung emission is $-0.1$, we thus set the uniform prior for this source to an allowed range between $-1$ and $+0.1$. The parameter $\beta$ is fixed in this source due to the limited number of measurements and to the degeneracy between $\beta$ and dust temperature. We choose a value consistent with those found in the literature for quasars and infrared bright galaxies \cite{beel06, kirk12,bend25}. We find that the parameter distributions do not change if we fix $\beta$ to higher or lower values within the expected 1 to 2.5 range, other than $<1\sigma$ changes in the dust temperature. The results are shown in Figure~\ref{fig:J1549ASED}.

We investigated the nature of J1549B using the full near-IR through radio spectral energy distribution (Figure~\ref{fig:J1549SEDnuLnu}).
Here we compare the flux densities to starburst templates from \cite{riek09}.
In both sources the F$_{\rm IR}$ luminosity on the Rayleigh-Jeans tail of the dust emission is likely dominated by thermal emission from cold dust, re-radiated from star formation activity. 
This requires a relatively modest level of 10$^{12}~L_\odot$ (3.8$\times10^{45}$~erg/s) worth of star formation luminosity. 
Using the \cite{bell03} calibrations between total IR luminosity and star formation rate (SFR) indicates a rate of 177 M${_\odot}$/year, which, for an AGN host at this redshift is a perfectly reasonable value. 
However, neither the amplitude of mid-infrared emission nor the radio emission can be explained by a starburst of this rate. 
In fact, even if the far-IR emission was best described by the 10$^{13}$~L${_\odot}$ template (indicating even larger SFR), the associated radio and mid-IR fluxes still could not explain our observations. 
Both sources have radio/IR fluxes that are $\sim$2 orders of magnitude above the level expected from SF, whereas the radio-FIR correlation in the SF galaxies is really quite tight. 
This indicates that J1549B likely hosts an AGN, boosting its near-IR and radio fluxes as compared to those expected purely from star formation.

The normalization to the \cite{riek09} template agrees with the value of L$_\mathrm{IR}$ that we infer from the results of the MCMC model fitting to the radio through submillimeter SED for J1549B. 
We find that the median, 16th and 84th percentile of the $log$(L$_{\rm IR}/L_\odot)={12.0 \pm 0.9}$ or 3.8$\times$10$^{45}$~erg/s, which indicates a total SFR of $(175_{-153}^{+1086}~M_\odot/yr$.
This places these quantities for J1549B amongst those that we expect for luminous infrared-selected AGN \cite{kirk19}.

\section{Molecular gas}
\label{sec:molecgas}

As part of a program to search for molecular gas outflows in $z\sim2$ quasars, \cite{bisc21} detail the detection of CO(4--3) emission in the J1549+1245 system. They detect J1549A and two very nearby ($<$20~kpc) companions, the second of which (in addition to the already introduced J1549B) -- J1549C -- is an emission line only source. 
These band 4 ALMA observations determined J1549A to be at $z_{\rm CO}$=2.374, distinctly different from the redshift determined by ionized gas emission in the optical/NIR observations. 
These data place J1549B at $z_{\rm CO}$=2.363 and J1549C at the same redshift as A, $z_{\rm CO}$=2.374. 

In the ALMA band 3 ($\sim$100~GHz) data we measure CO(3--2) emission and in the VLA Ka ($\sim$30~GHz) data we measure CO(1--0) emission in J1549A and J1549B. We do not detect J1549C in either of the other CO transitions.
We extract emission line information from continuum-subtracted data cubes. 
We select the spectral windows containing and bracketing the CO emission lines, and use the CASA module \textsc{uvcontsub} to fit the continuum spectrum on either side of the lines using a zeroth order polynomial. 
This model fits a line to the millimeter-wave continuum emission and then subtracts it from the entire spectral cube, leaving only the emission from the CO emission lines. 
We image the cubes using \textsc{tclean} with natural weighting and automasking. The CO(1--0) was imaged at 100~km/s channel widths, and the CO(3--2) was imaged at 45~km/s channel widths. 
We follow the procedure of \citep{bisc21} and image the CO(4--3) at 40~km/s channel widths. 

\begin{figure}[!htbp]
\centering
    \includegraphics[width=\columnwidth]{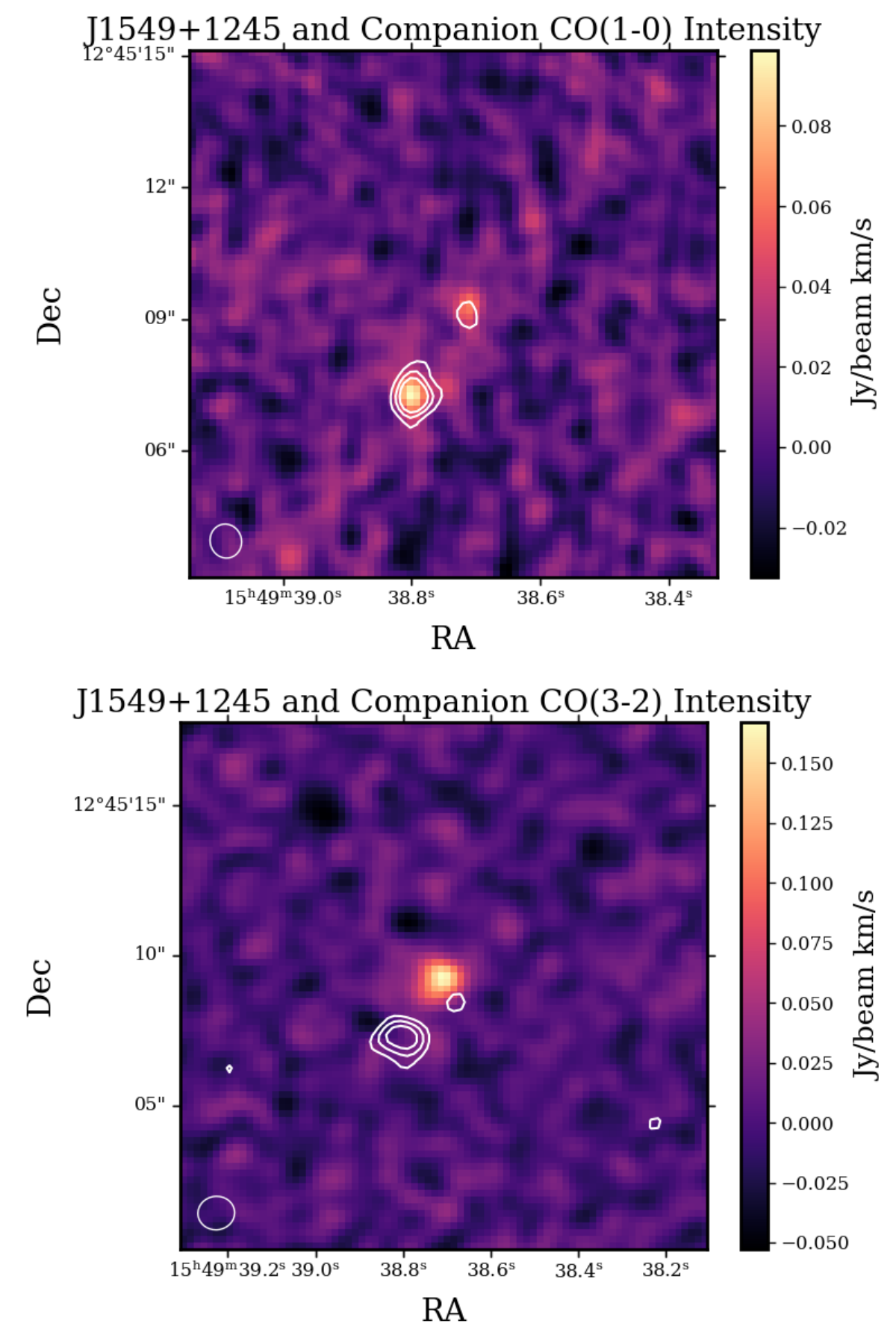}
    \caption{Carbon monoxide (CO) intensities for the J1549+1245 system. The top panel is CO(1--0) and the bottom panel is CO(3--2). The shading is the result of selecting the spectral slabs containing the CO emission for J1549A, while the white contours overlay the CO emission from the spectral slabs containing J1549B. Note that in the case of the CO(1--0) emission, we cannot spectrally isolate the two sources. The beam sizes are shown as white ellipses in the lower left corner and are X"xY" and 1.3"x1.2" for the (1--0) and (3--2) maps, respectively.}
    \label{fig:J1549CO32_int}
\end{figure}

\begin{figure*}[!htbp]
\centering
    \includegraphics[width=\textwidth]{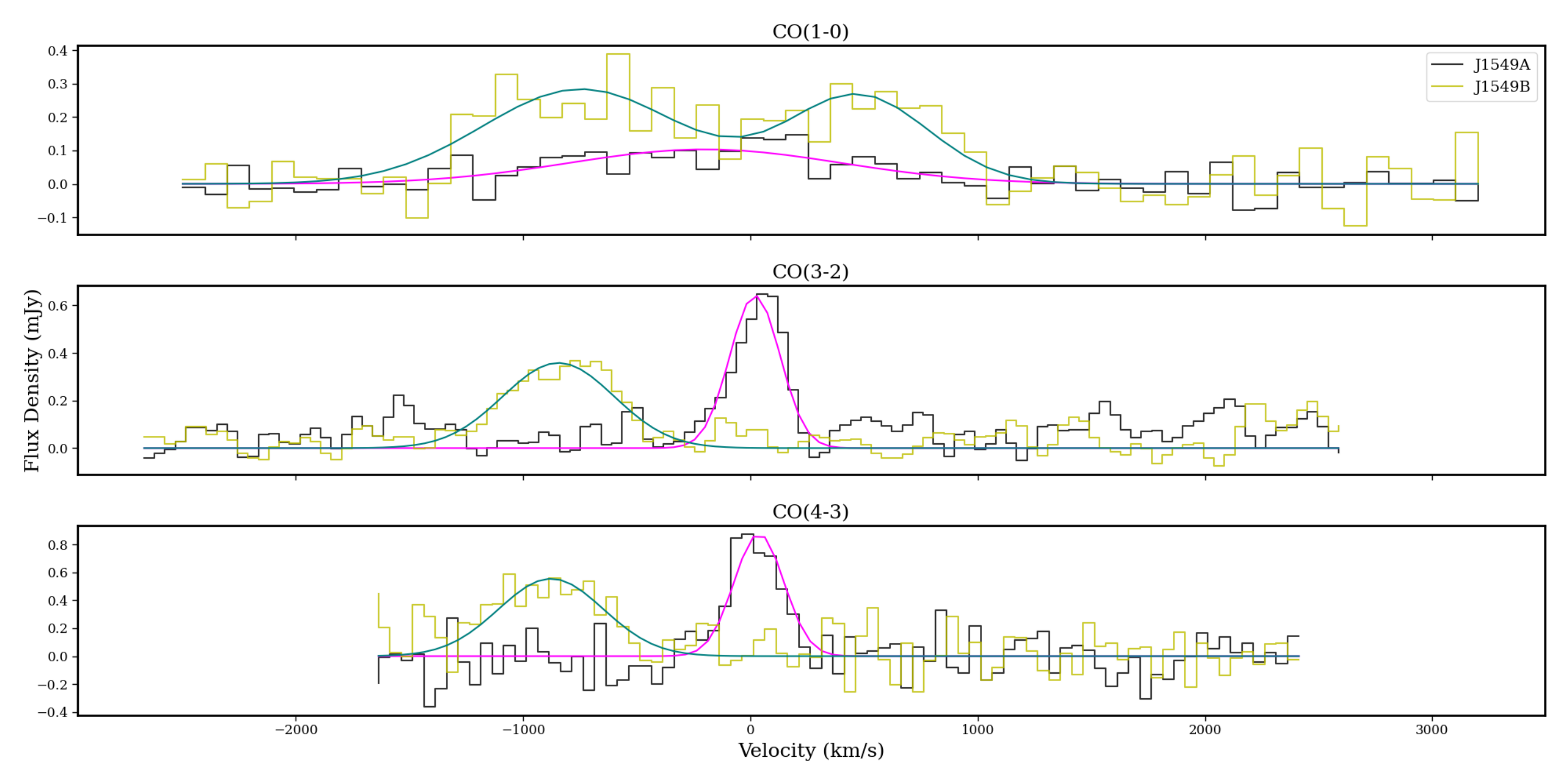}
 \caption{Spectra of the three CO lines for the J1549+1245 system. J1549A measured spectrum is in black and it's best-fitting Gaussian is in magenta, while those for J1549B are shown in yellow and teal, respectively. The transitions are indicated in each subplot title.}
    \label{fig:J1549COall}
\end{figure*}

The total intensity maps of the  CO(1--0) and CO(3--2) emission are shown in Figure~\ref{fig:J1549CO32_int} left and right, respectively. 
The physical and spectral distribution of CO(3--2) in J1549A and J1549B are distinct, such that we can separate the total intensities (Figure~\ref{fig:J1549CO32_int}) and velocity line profiles (Figure~\ref{fig:J1549COall}).
The spectral extent of the CO(1--0) emission is distinct from the other transitions; it is very broad and in the case of J1549B it is double peaked, with the second peak redshifted with respect to J1549A's systemic velocity (Figure~\ref{fig:J1549COall}).  

We also examine the implications of the three CO transitions detected in both J1549A and J1549B. 
Figure~\ref{fig:J1549COall} shows the velocity profiles for all three energy transitions for both sources with the x-axis scaled to the systemic velocity of J1549A. 
CO(3--2) and CO(4--3) are narrower lines and thus provide more accurate redshift measurements for the host galaxies. We fit these lines with single Gaussian curves and confirm the redshift of J1549A that is reported in \cite{bisc21}, z$_\mathrm{CO(3-2),A}$=2.374, but we find z$_\mathrm{CO(3-2),B}$=2.365, each with an uncertainty of $\Delta$z=0.001. The value reported in \cite{bisc21} for J1549B is z$_\mathrm{CO(3-2),B}$=2.363$\pm$0.001. 
The CO(1--0) line is significantly broader for both sources, and in the case of J1549B is better fit by two Gaussian functions as it has a more double-peaked shape with the redshifted peak centered at (-762$\pm$71)~km/s relative to the systemic velocity of J1549A and the blueshifted peak at (438$\pm$72)~km/s. This shape of the line profile indicates the presence of CO gas from J1549B that spectrally overlaps with J1549A, which may indicate merger activity. If the CO(1--0) gas is optically thinner than the higher energy transitions, then we may be tracing the full rotation of a disk in J1549B. 
We note that \cite{bisc21} also report the presence of a second companion, J1549C, as seen in CO(4--3) emission only. We observe a possible 2$\sigma$ CO(3--2) line detection of this source, but do not observe it in CO(1--0). Thus, the existence of another companion remains ambiguous.
 
We compute the CO line luminosity from the CO(3--2) measurement for J1549A and J1549B as:
\begin{equation}
    \mathrm{L}^\prime_{\rm CO} = 3.25\times 10^{7} \mathrm{S}_{\rm CO} \Delta \mathrm{v} \frac{ D_{\rm L}^2}{\nu^3_{\rm obs} (1+z)^3}\,  \rm K\, km\, s^{-1}\, pc^2,
\end{equation}
where S$_\mathrm{CO} \Delta$v is the integrated line flux, $\nu_{obs}$ is the observed frequency of the CO emission line at a given redshift, and $D_L$ is the luminosity distance \citep{cari13}. 
L$^\prime_\mathrm{CO}$ computed from the CO(3--2) lines yields to the same value within uncertainty found by \cite{bisc17} using the CO(4--3) line: L$^\prime_\mathrm{CO(3-2), A}$ = (0.55$\pm$0.07)$\times$10$^{10}$~K~km~s$^{-1}$, L$^\prime_\mathrm{CO(3-2), B}$ = (0.67$\pm$0.04)$\times$10$^{10}$~K~km~s$^{-1}$. L$^\prime_\mathrm{CO}$ from the CO(1--0) line is comparable for J1549A with L$^\prime_\mathrm{CO(3-2), A}$/L$^\prime_\mathrm{CO(1-0), A}$ = 1.15. 
The CO(1--0) line luminosity for J1549B, L$^\prime_\mathrm{CO(1-0), B}$ is (1.57$\pm$0.05)$\times$10$^{10}$~K~km~s$^{-1}$. 
This value is much higher than the others due to the width of the line. 
Several recent studies including \cite{bisc21}, \cite{pern18}, and \cite{kirk19} report on the L$^\prime_\mathrm{CO}$--L$_\mathrm{IR}$ relation for obscured and unobscured quasars in comparison to main sequence galaxies and submillimeter galaxies.
The analysis carried out in \cite{kirk19} considers only CO-detected sources and produced median CO spectral line energy distributions (SLEDs) for obscured quasars in comparison to the full sample used in \cite{cari13}, and match the objects according to IR luminosity, redshift, and lensing magnitude. 
The analyses carried out in \cite{bisc21} and \cite{pern18} find that high-z quasars are fainter in L$^\prime_\mathrm{CO}$ for a given L$_\mathrm{IR}$ in comparison to other galaxies, while \cite{kirk19} report no significant difference. The latter study uses diverse SLEDs and ones that are more similar to submillimeter galaxies, which could explain the difference.
Since we now have direct measurements of L$^\prime_\mathrm{CO(1-0)}$ for both J1549A and J1549B, we can add new results to this discussion for these two sources. 
We find that J1549A lies below the main sequence and SMGs on the L$^\prime_\mathrm{CO}$ vs L$_{\rm IR}$ plot, similar to the estimation provided by \cite{bisc21}. For J1549B, however, we find that it sits just on the lower edge of the standard deviation of the L$^\prime_\mathrm{CO}$--L$_\mathrm{IR}$ relation for normal galaxies as determined by \cite{sarg14}, for which SMGs also agree.
This supports the notion that J1549B is a highly dusty star-forming galaxy, with excess radio and near-IR emissions that are consistent with quasar activity.


\section{The origin of radio emission and other possible mechanisms for the SED shapes}
\label{alternativemodels}

The physical mechanisms driving the radio emission in the population of radio quiet quasars remains a matter of exploration. 
Radio emission may be a result of scaled-down relativistic jets, quasar winds, coronal emission from the accretion disk, free-free emission from the quasar radiation photo-ionizing ambient gas, star formation activity driving synchrotron radiation or free-free emission, or combinations of any of the above \citep{zaka14, pane19}. 
Some studies show that the spectral shape and area of origin of the radio emission correlates with the Eddington ratio \citep{alho22, chen25}, indicating that careful assessment of interferometric data from varying scales is necessary to best understand the emission mechanisms. 
At the physical scale of the VLA data used in this analysis ($\sim$8~kpc), it is not possible to differentiate the relative contributions to the total radio emission from accretion disk scales versus larger scale emission.
Further complicating the interpretation, the number of radio quiet quasars with tens to hundreds of GHz observations is limited, with recent studies \citep{bald22} targeting their expansion at high angular resolution ($\sim$50~mas). 
The majority of quasar-driven radio emission mechanisms indicate spectra consistent with flat or power law slopes \cite{pane19}, while starburst galaxies may exhibit exponential cutoffs in their non-thermal components \citep{klei18}. 

Here, we report on our tests of alternative models to fit our observed SED of the radio through sub-mm emission of J1549B.
We observe a steep spectrum source ($\alpha_{synchrotron} \leq$-0.5) in the radio indicative of optically thin synchrotron emission, and explore models that could physically produce this.
The first is using only a power law synchrotron spectrum and a modified blackbody dust spectrum, without the spectral distortion due to the tSZ. 
The data cannot be well fit within reasonable parameter limits (namely, 1$\leq \beta \leq$2.5), and produces a reduced $\chi^2_{\rm red}$ of 27. 
A double power law synchrotron model in which there is a spectral break to steeper spectral index after some frequency \cite{poll00, pane19, azad23} that we allow to vary also fails to fit the data with a reduced $\chi^2_{\rm red}$ of 27. 
A model that includes both a double power law and the tSZ spectral distortion produces a fit with a reduced $\chi^2_{\rm red}$ of 18. 
This model has one additional parameter, leaving only two degrees of freedom, whereas our fiducial model has three degrees of freedom. 
The double power law model still prefers the inclusion of the tSZ spectral distortion at the 4.6$\sigma$ level. 

We also fit the data with a synchrotron model with an exponential cutoff at a free-floating frequency, $\nu_\mathrm{cutoff}>$0~GHz, $S_\nu \propto \nu^\alpha e^{-\nu/\nu_\mathrm{cutoff}}$.
We run an MCMC using \textsc{emcee} as in our fiducial model, with a flat, broad prior on (observed) cutoff frequency, [1e8, 1e14]~Hz, and the other priors the same as the fiducial model. 
We find a parameter space that fits these data using a combination of an exponentially cutoff power law and a modified blackbody dust emission.
We include contour plots of the posterior distributions in the Extended Data (Figure~\ref{fig:J1549B_corner_expPL}).
While the reduced $\chi^2$ of the model curve produced from the medians of the parameter posterior distributions is worse than our fiducial model, $\chi^2_\mathrm{red}=14.8$, we perform a k-fold cross-validation routine to estimate the Expected Predictive Log Density (ELPD).
We run this routine on both the fiducial tSZ model and the exponentially cutoff synchrotron model, and find a preference for the tSZ model, with $\Delta$ELPD$_\mathrm{tSZ-expPL}=17.6$ with a 95\% confidence interval of $\pm$5.5. 
We also ran each model using \textsc{dynesty} \citep{spea20,kopo25,higs19,fero09,neal03,hand15a,hand15b} in order to compute the Bayesian evidences.
We find the difference in the logarithm of the evidence between the tSZ model and the exponential power law model to be $\Delta$ln$Z=$1.03, with effective significance $\sigma_\mathrm{eff}$=1.44. 
The log-evidences do not provide a statistically significant indicator for which model is best, though this metric tends to be heavily dependent on priors.  
Crucially, the resulting parameters from the exponentially cutoff synchrotron emission pose some challenges for the physical interpretation of the underlying emission spectra. 
With respect to the dust, this model requires a steep dust spectrum with emissivity index of 2.1 and total infrared luminosity of 1.7$\times$10$^{13}$~L$_\odot$ or 6.5$\times$10$^{46}$~ergs/s. 
The reason for this is evident: the emission spectrum from the dust requires the modified blackbody increase in the sub-mm regime (the upturn in the spectrum at frequencies $>$ 150~GHz). 
It is also evident that the model dust spectrum on its own cannot account for the curved distortion indicated by the data between 150 and 250 GHz, which is more naturally reproduced by the spectral distortion from the tSZ.  
We explored the possibility of a two-temperature dust model to fit this curvature by requiring the sum of a cold (T1 [10~K,45~K]) and a warm (T2 [45~K, 100~K]) modified blackbody SED, where beta is free to vary as in the fiducial model for each component. These data cannot constrain this model well, but we find it would require the warm and cold components to have emissivities that are different by $\Delta \beta \geq 0.5$. We can also see this necessity by simply examining the slope of the fluxes between 95 and 230~GHz. Such a model implies very different dust properties of each component and is not physically justified. 
We also performed a grid search on the value of $\beta$ by fixing it to different values between 1.5 and 2.5. We found the significance of the tSZ amplitude varies between 6 and 6.7 sigma. 
The amplitude itself varies by 1 sigma between the extreme ends of the beta range. 
While a multi-temperature dust spectrum cannot be statistically ruled out with the current frequency sampling, our preference for the single-temperature plus tSZ model is based on both its statistical performance relative to alternatives we could test and on the physical implausibility of the parameter values required by the alternatives we could constrain. 

For the exponentially cutoff synchrotron model, the rest frequency of the exponential cutoff is 176$^{+31}_{-27}$~GHz. 
This high frequency cutoff has only been seen in extreme radio-loud AGN \cite{poll00,ragi16,azad23} with radio luminosities two orders of magnitude above that of our target. 
Following \cite{ragi16}, we consider the possibility that this spectrum arises from synchrotron radiation from a powerful compact (1-5~pc) radio source.  
For a model with the total luminosity and rest frame cutoff frequency fixed to that of our target, we compute the expected photon energy density (infrared irradiated and CMB radiation separately), synchrotron energy density, and magnetic energy density for a range of magnetic field strengths as well as the Synchrotron Self Compton cooling times.  
We include in this calculation the consideration of inverse Compton cooling. 
The CMB radiation energy density is one to five orders of magnitude smaller than the other energy density components (IR radiative, magnetic, and synchrotron). 
We find that for a wide range of assumed magnetic field strengths the synchrotron self-Compton cooling times are very short making it impossible to sustain electrons with sufficient energy to produce the 176~GHz cutoff.
Equipartition requires an unfeasible magnetic field strength 0.1 G, or else an emitting region that is $\sim$160 pc, which is also highly unlikely for this type of source. 
Quasar winds are expected to extend this far (and beyond), but to produce the high cutoff frequency we observe requires a strong magnetic field out to these relatively large distances and/or a very slow wind speed ($\leq0.01c$), which does not align with the other observations of this system. 

A weak or mini jet with the same radio power can have different extents and shapes – flat or steep; though most in the literature observed with VLBI/VLBA have flat spectra at higher frequencies as the emission becomes optically thick towards the base/core, which would also cause the spectral flattening at high frequencies \cite{bald22}. 
Crucially, the theoretical scenario that allows for a jet of emission to cause the exponential cutoff at 176~GHz requires a different set of parameters. The jet scenario faces similar challenges as the wind scenario in that the estimated magnetic field strength decreases and the escape time exceeds that of cooling time. This would either not produce the break at all for the energy range of interest or else would produce the high energy steepening of alpha by $\sim$0.5 (a double power law model, which we find to be statistically inferior as it produces a worse fit to the data with even fewer degrees of freedom).

While the synchrotron emission in the radio regime is almost certainly due to one or a combination of these physical mechanisms, none of them can plausibly produce such a high frequency exponential cutoff in the emission at rest frame 176~GHz, given what we know about the system. 
Adding very long baseline observations of this system at observed 22 and 45~GHz would certainly help to illuminate the source of the radio emission.
Moreover, the SED exhibits a spectral distortion in the synchrotron dominated portion of the spectrum, but also in the shape of the dust spectrum. 
As such, our tSZ model is favored as both the most statistically sound and physically favored model.






\bibliographystyle{naturea}
\bibliography{kh_master}

\begin{thebibliography}{10}
\expandafter\ifx\csname url\endcsname\relax
  \def\url#1{\burl{#1}}\fi
\expandafter\ifx\csname urlprefix\endcsname\relax\def\urlprefix{URL }\fi
\providecommand{\bibinfo}[2]{#2}
\providecommand{\eprint}[2][]{\url{#2}}
\providecommand{\doi}[1]{\url{https://doi.org/#1}}

\bibitem{mago98}
\bibinfo{author}{{Magorrian}, J.} \emph{et~al.}
\newblock \bibinfo{title}{{The Demography of Massive Dark Objects in Galaxy Centers}}.
\newblock \emph{\bibinfo{journal}{\aj}} \textbf{\bibinfo{volume}{115}}, \bibinfo{pages}{2285--2305} (\bibinfo{year}{1998}).

\bibitem{gebh00}
\bibinfo{author}{{Gebhardt}, K.} \emph{et~al.}
\newblock \bibinfo{title}{{A Relationship between Nuclear Black Hole Mass and Galaxy Velocity Dispersion}}.
\newblock \emph{\bibinfo{journal}{\apjl}} \textbf{\bibinfo{volume}{539}}, \bibinfo{pages}{L13--L16} (\bibinfo{year}{2000}).

\bibitem{ferr00}
\bibinfo{author}{{Ferrarese}, L.} \& \bibinfo{author}{{Merritt}, D.}
\newblock \bibinfo{title}{{A Fundamental Relation between Supermassive Black Holes and Their Host Galaxies}}.
\newblock \emph{\bibinfo{journal}{\apjl}} \textbf{\bibinfo{volume}{539}}, \bibinfo{pages}{L9--L12} (\bibinfo{year}{2000}).

\bibitem{ruan15}
\bibinfo{author}{{Ruan}, J.~J.}, \bibinfo{author}{{McQuinn}, M.} \& \bibinfo{author}{{Anderson}, S.~F.}
\newblock \bibinfo{title}{{Detection of Quasar Feedback from the Thermal Sunyaev-Zel{\textquotesingle}dovich Effect in Planck}}.
\newblock \emph{\bibinfo{journal}{\apj}} \textbf{\bibinfo{volume}{802}}, \bibinfo{pages}{135} (\bibinfo{year}{2015}).

\bibitem{verd16}
\bibinfo{author}{{Verdier}, L.} \emph{et~al.}
\newblock \bibinfo{title}{{Quasar host environments: The view from Planck}}.
\newblock \emph{\bibinfo{journal}{\aap}} \textbf{\bibinfo{volume}{588}}, \bibinfo{pages}{A61} (\bibinfo{year}{2016}).

\bibitem{cric16}
\bibinfo{author}{{Crichton}, D.} \emph{et~al.}
\newblock \bibinfo{title}{{Evidence for the thermal Sunyaev-Zel'dovich effect associated with quasar feedback}}.
\newblock \emph{\bibinfo{journal}{\mnras}} \textbf{\bibinfo{volume}{458}}, \bibinfo{pages}{1478--1492} (\bibinfo{year}{2016}).

\bibitem{soer17}
\bibinfo{author}{{Soergel}, B.}, \bibinfo{author}{{Giannantonio}, T.}, \bibinfo{author}{{Efstathiou}, G.}, \bibinfo{author}{{Puchwein}, E.} \& \bibinfo{author}{{Sijacki}, D.}
\newblock \bibinfo{title}{{Constraints on AGN feedback from its Sunyaev-Zel'dovich imprint on the cosmic background radiation}}.
\newblock \emph{\bibinfo{journal}{\mnras}} \textbf{\bibinfo{volume}{468}}, \bibinfo{pages}{577--596} (\bibinfo{year}{2017}).

\bibitem{lacy19}
\bibinfo{author}{{Lacy}, M.} \emph{et~al.}
\newblock \bibinfo{title}{{Direct detection of quasar feedback via the Sunyaev-Zeldovich effect}}.
\newblock \emph{\bibinfo{journal}{\mnras}} \textbf{\bibinfo{volume}{483}}, \bibinfo{pages}{L22--L27} (\bibinfo{year}{2019}).

\bibitem{hall19}
\bibinfo{author}{{Hall}, K.~R.} \emph{et~al.}
\newblock \bibinfo{title}{{Quantifying the thermal Sunyaev-Zel'dovich effect and excess millimetre emission in quasar environments}}.
\newblock \emph{\bibinfo{journal}{\mnras}} \textbf{\bibinfo{volume}{490}}, \bibinfo{pages}{2315--2335} (\bibinfo{year}{2019}).

\bibitem{brow19}
\bibinfo{author}{{Brownson}, S.}, \bibinfo{author}{{Maiolino}, R.}, \bibinfo{author}{{Tazzari}, M.}, \bibinfo{author}{{Carniani}, S.} \& \bibinfo{author}{{Henden}, N.}
\newblock \bibinfo{title}{{Detecting the halo heating from AGN feedback with ALMA}}.
\newblock \emph{\bibinfo{journal}{\mnras}} \textbf{\bibinfo{volume}{490}}, \bibinfo{pages}{5134--5146} (\bibinfo{year}{2019}).

\bibitem{suny70}
\bibinfo{author}{{Sunyaev}, R.~A.} \& \bibinfo{author}{{Zeldovich}, Y.~B.}
\newblock \bibinfo{title}{{The Spectrum of Primordial Radiation, its Distortions and their Significance}}.
\newblock \emph{\bibinfo{journal}{Comments on Astrophys. Space Phys.}} \textbf{\bibinfo{volume}{2}}, \bibinfo{pages}{66} (\bibinfo{year}{1970}).

\bibitem{fauc12}
\bibinfo{author}{{Faucher-Gigu{\`e}re}, C.-A.} \& \bibinfo{author}{{Quataert}, E.}
\newblock \bibinfo{title}{{The physics of galactic winds driven by active galactic nuclei}}.
\newblock \emph{\bibinfo{journal}{\mnras}} \textbf{\bibinfo{volume}{425}}, \bibinfo{pages}{605--622} (\bibinfo{year}{2012}).

\bibitem{zubo12}
\bibinfo{author}{{Zubovas}, K.} \& \bibinfo{author}{{King}, A.}
\newblock \bibinfo{title}{{Clearing Out a Galaxy}}.
\newblock \emph{\bibinfo{journal}{\apjl}} \textbf{\bibinfo{volume}{745}}, \bibinfo{pages}{L34} (\bibinfo{year}{2012}).

\bibitem{gill17}
\bibinfo{author}{{Gilli}, R.}, \bibinfo{author}{{Calura}, F.}, \bibinfo{author}{{D'Ercole}, A.} \& \bibinfo{author}{{Norman}, C.}
\newblock \bibinfo{title}{{Exponentially growing bubbles around early supermassive black holes}}.
\newblock \emph{\bibinfo{journal}{\aap}} \textbf{\bibinfo{volume}{603}}, \bibinfo{pages}{A69} (\bibinfo{year}{2017}).

\bibitem{hadz25a}
\bibinfo{author}{{Hadzhiyska}, B.}, \bibinfo{author}{{Ferraro}, S.} \& \bibinfo{author}{{Zhou}, R.}
\newblock \bibinfo{title}{{Tracing cosmic gas in filaments and halos: Low-redshift insights from the kinematic Sunyaev-Zel'dovich effect}}.
\newblock \emph{\bibinfo{journal}{\prd}} \textbf{\bibinfo{volume}{111}}, \bibinfo{pages}{023534} (\bibinfo{year}{2025}).

\bibitem{pand25}
\bibinfo{author}{{Pandey}, S.} \emph{et~al.}
\newblock \bibinfo{title}{{Constraints on cosmology and baryonic feedback with joint analysis of Dark Energy Survey Year 3 lensing data and ACT DR6 thermal Sunyaev-Zel'dovich effect observations}}.
\newblock \emph{\bibinfo{journal}{arXiv e-prints}} \bibinfo{pages}{arXiv:2506.07432} (\bibinfo{year}{2025}).

\bibitem{hadz25b}
\bibinfo{author}{{Hadzhiyska}, B.} \emph{et~al.}
\newblock \bibinfo{title}{{Evidence for large baryonic feedback at low and intermediate redshifts from kinematic Sunyaev-Zel'dovich observations with ACT and DESI photometric galaxies}}.
\newblock \emph{\bibinfo{journal}{\prd}} \textbf{\bibinfo{volume}{112}}, \bibinfo{pages}{083509} (\bibinfo{year}{2025}).

\bibitem{ried25}
\bibinfo{author}{{Ried Guachalla}, B.} \emph{et~al.}
\newblock \bibinfo{title}{{Backlighting extended gas halos around luminous red galaxies: Kinematic Sunyaev-Zel'dovich effect from DESI Y1 and ACT data}}.
\newblock \emph{\bibinfo{journal}{\prd}} \textbf{\bibinfo{volume}{112}}, \bibinfo{pages}{103512} (\bibinfo{year}{2025}).

\bibitem{mroc19}
\bibinfo{author}{{Mroczkowski}, T.} \emph{et~al.}
\newblock \bibinfo{title}{{Astrophysics with the Spatially and Spectrally Resolved Sunyaev-Zeldovich Effects. A Millimetre/Submillimetre Probe of the Warm and Hot Universe}}.
\newblock \emph{\bibinfo{journal}{\ssr}} \textbf{\bibinfo{volume}{215}}, \bibinfo{pages}{17} (\bibinfo{year}{2019}).

\bibitem{goba19}
\bibinfo{author}{{Gobat}, R.} \emph{et~al.}
\newblock \bibinfo{title}{{Sunyaev-Zel'dovich detection of the galaxy cluster Cl J1449+0856 at z = 1.99: The pressure profile in uv space}}.
\newblock \emph{\bibinfo{journal}{\aap}} \textbf{\bibinfo{volume}{629}}, \bibinfo{pages}{A104} (\bibinfo{year}{2019}).

\bibitem{dima23}
\bibinfo{author}{{Di Mascolo}, L.} \emph{et~al.}
\newblock \bibinfo{title}{{Forming intracluster gas in a galaxy protocluster at a redshift of 2.16}}.
\newblock \emph{\bibinfo{journal}{\nat}} \textbf{\bibinfo{volume}{615}}, \bibinfo{pages}{809--812} (\bibinfo{year}{2023}).

\bibitem{suny72}
\bibinfo{author}{{Sunyaev}, R.~A.} \& \bibinfo{author}{{Zeldovich}, Y.~B.}
\newblock \bibinfo{title}{{Formation of Clusters of Galaxies; Protocluster Fragmentation and Intergalactic Gas Heating}}.
\newblock \emph{\bibinfo{journal}{\aap}} \textbf{\bibinfo{volume}{20}}, \bibinfo{pages}{189} (\bibinfo{year}{1972}).

\bibitem{cari98}
\bibinfo{author}{{Carilli}, C.~L.} \emph{et~al.}
\newblock \bibinfo{title}{{An X-Ray Cluster at Redshift 2.156?}}
\newblock \emph{\bibinfo{journal}{\apjl}} \textbf{\bibinfo{volume}{494}}, \bibinfo{pages}{L143--L146} (\bibinfo{year}{1998}).

\bibitem{over05}
\bibinfo{author}{{Overzier}, R.~A.} \emph{et~al.}
\newblock \bibinfo{title}{{On the X-ray emission of z {\ensuremath{\sim}}2 radio galaxies: IC scattering of the CMB and no evidence for fully-formed potential wells}}.
\newblock \emph{\bibinfo{journal}{\aap}} \textbf{\bibinfo{volume}{433}}, \bibinfo{pages}{87--100} (\bibinfo{year}{2005}).

\bibitem{cham21}
\bibinfo{author}{{Champagne}, J.~B.} \emph{et~al.}
\newblock \bibinfo{title}{{Comprehensive Gas Characterization of a z = 2.5 Protocluster: A Cluster Core Caught in the Beginning of Virialization?}}
\newblock \emph{\bibinfo{journal}{\apj}} \textbf{\bibinfo{volume}{913}}, \bibinfo{pages}{110} (\bibinfo{year}{2021}).

\bibitem{tozz22}
\bibinfo{author}{{Tozzi}, P.} \emph{et~al.}
\newblock \bibinfo{title}{{The 700 ks Chandra Spiderweb Field. II. Evidence for inverse-Compton and thermal diffuse emission in the Spiderweb galaxy}}.
\newblock \emph{\bibinfo{journal}{\aap}} \textbf{\bibinfo{volume}{667}}, \bibinfo{pages}{A134} (\bibinfo{year}{2022}).

\bibitem{jone23}
\bibinfo{author}{{Jones}, G.~C.} \emph{et~al.}
\newblock \bibinfo{title}{{An investigation of the circumgalactic medium around z {\ensuremath{\sim}} 2.2 AGN with ACA and ALMA}}.
\newblock \emph{\bibinfo{journal}{\mnras}} \textbf{\bibinfo{volume}{522}}, \bibinfo{pages}{275--291} (\bibinfo{year}{2023}).

\bibitem{Benn25}
\bibinfo{author}{{Bennett}, J.~S.} \emph{et~al.}
\newblock \bibinfo{title}{{Prevention is better than cure? Feedback from high specific energy winds in cosmological simulations with ARKENSTONE}}.
\newblock \emph{\bibinfo{journal}{\mnras}} \textbf{\bibinfo{volume}{543}}, \bibinfo{pages}{1456--1478} (\bibinfo{year}{2025}).

\bibitem{plan20}
\bibinfo{author}{{Planck Collaboration}} \emph{et~al.}
\newblock \bibinfo{title}{{Planck 2018 results. VI. Cosmological parameters}}.
\newblock \emph{\bibinfo{journal}{\aap}} \textbf{\bibinfo{volume}{641}}, \bibinfo{pages}{A6} (\bibinfo{year}{2020}).

\bibitem{bisc17}
\bibinfo{author}{{Bischetti}, M.} \emph{et~al.}
\newblock \bibinfo{title}{{The WISSH quasars project. I. Powerful ionised outflows in hyper-luminous quasars}}.
\newblock \emph{\bibinfo{journal}{\aap}} \textbf{\bibinfo{volume}{598}}, \bibinfo{pages}{A122} (\bibinfo{year}{2017}).

\bibitem{bisc21}
\bibinfo{author}{{Bischetti}, M.} \emph{et~al.}
\newblock \bibinfo{title}{{The WISSH quasars project. IX. Cold gas content and environment of luminous QSOs at z {\ensuremath{\sim}} 2.4-4.7}}.
\newblock \emph{\bibinfo{journal}{\aap}} \textbf{\bibinfo{volume}{645}}, \bibinfo{pages}{A33} (\bibinfo{year}{2021}).

\bibitem{viet20}
\bibinfo{author}{{Vietri}, G.} \emph{et~al.}
\newblock \bibinfo{title}{{SUPER. III. Broad line region properties of AGNs at z {\ensuremath{\sim}} 2}}.
\newblock \emph{\bibinfo{journal}{\aap}} \textbf{\bibinfo{volume}{644}}, \bibinfo{pages}{A175} (\bibinfo{year}{2020}).

\bibitem{bell03}
\bibinfo{author}{{Bell}, E.~F.}
\newblock \bibinfo{title}{{Estimating Star Formation Rates from Infrared and Radio Luminosities: The Origin of the Radio-Infrared Correlation}}.
\newblock \emph{\bibinfo{journal}{\apj}} \textbf{\bibinfo{volume}{586}}, \bibinfo{pages}{794--813} (\bibinfo{year}{2003}).

\bibitem{zaka16}
\bibinfo{author}{{Zakamska}, N.~L.} \emph{et~al.}
\newblock \bibinfo{title}{{Star formation in quasar hosts and the origin of radio emission in radio-quiet quasars}}.
\newblock \emph{\bibinfo{journal}{\mnras}} \textbf{\bibinfo{volume}{455}}, \bibinfo{pages}{4191--4211} (\bibinfo{year}{2016}).

\bibitem{hwan18}
\bibinfo{author}{{Hwang}, H.-C.} \emph{et~al.}
\newblock \bibinfo{title}{{Winds as the origin of radio emission in z = 2.5 radio-quiet extremely red quasars}}.
\newblock \emph{\bibinfo{journal}{\mnras}} \textbf{\bibinfo{volume}{477}}, \bibinfo{pages}{830--844} (\bibinfo{year}{2018}).

\bibitem{laor19}
\bibinfo{author}{{Laor}, A.}, \bibinfo{author}{{Baldi}, R.~D.} \& \bibinfo{author}{{Behar}, E.}
\newblock \bibinfo{title}{{What drives the radio slopes in radio-quiet quasars?}}
\newblock \emph{\bibinfo{journal}{\mnras}} \textbf{\bibinfo{volume}{482}}, \bibinfo{pages}{5513--5523} (\bibinfo{year}{2019}).

\bibitem{pane19}
\bibinfo{author}{{Panessa}, F.} \emph{et~al.}
\newblock \bibinfo{title}{{The origin of radio emission from radio-quiet active galactic nuclei}}.
\newblock \emph{\bibinfo{journal}{Nat. Astron.}} \textbf{\bibinfo{volume}{3}}, \bibinfo{pages}{387--396} (\bibinfo{year}{2019}).

\bibitem{dey19}
\bibinfo{author}{{Dey}, A.} \emph{et~al.}
\newblock \bibinfo{title}{{Overview of the DESI Legacy Imaging Surveys}}.
\newblock \emph{\bibinfo{journal}{Astron. J.}} \textbf{\bibinfo{volume}{157}}, \bibinfo{pages}{168} (\bibinfo{year}{2019}).

\bibitem{carl02}
\bibinfo{author}{{Carlstrom}, J.~E.}, \bibinfo{author}{{Holder}, G.~P.} \& \bibinfo{author}{{Reese}, E.~D.}
\newblock \bibinfo{title}{{Cosmology with the Sunyaev-Zel'dovich Effect}}.
\newblock \emph{\bibinfo{journal}{\araa}} \textbf{\bibinfo{volume}{40}}, \bibinfo{pages}{643--680} (\bibinfo{year}{2002}).

\bibitem{rich06}
\bibinfo{author}{{Richards}, G.~T.} \emph{et~al.}
\newblock \bibinfo{title}{{Spectral Energy Distributions and Multiwavelength Selection of Type 1 Quasars}}.
\newblock \emph{\bibinfo{journal}{\apjs}} \textbf{\bibinfo{volume}{166}}, \bibinfo{pages}{470--497} (\bibinfo{year}{2006}).

\bibitem{cava10}
\bibinfo{author}{{Cavagnolo}, K.~W.} \emph{et~al.}
\newblock \bibinfo{title}{{A Relationship Between AGN Jet Power and Radio Power}}.
\newblock \emph{\bibinfo{journal}{\apj}} \textbf{\bibinfo{volume}{720}}, \bibinfo{pages}{1066--1072} (\bibinfo{year}{2010}).

\bibitem{harr24}
\bibinfo{author}{{Harrison}, C.~M.} \& \bibinfo{author}{{Ramos Almeida}, C.}
\newblock \bibinfo{title}{{Observational Tests of Active Galactic Nuclei Feedback: An Overview of Approaches and Interpretation}}.
\newblock \emph{\bibinfo{journal}{Galaxies}} \textbf{\bibinfo{volume}{12}}, \bibinfo{pages}{17} (\bibinfo{year}{2024}).

\bibitem{mcca87}
\bibinfo{author}{{McCarthy}, P.~J.}, \bibinfo{author}{{van Breugel}, W.} \& \bibinfo{author}{{Heckman}, T.}
\newblock \bibinfo{title}{{Evidence for Large-Scale Winds from Starburst Galaxies. I. The Nature of the Ionized Gas in M82 and NGC 253}}.
\newblock \emph{\bibinfo{journal}{\aj}} \textbf{\bibinfo{volume}{93}}, \bibinfo{pages}{264} (\bibinfo{year}{1987}).

\bibitem{stri00}
\bibinfo{author}{{Strickland}, D.~K.} \& \bibinfo{author}{{Stevens}, I.~R.}
\newblock \bibinfo{title}{{Starburst-driven galactic winds - I. Energetics and intrinsic X-ray emission}}.
\newblock \emph{\bibinfo{journal}{\mnras}} \textbf{\bibinfo{volume}{314}}, \bibinfo{pages}{511--545} (\bibinfo{year}{2000}).

\bibitem{hend18}
\bibinfo{author}{{Henden}, N.~A.}, \bibinfo{author}{{Puchwein}, E.}, \bibinfo{author}{{Shen}, S.} \& \bibinfo{author}{{Sijacki}, D.}
\newblock \bibinfo{title}{{The FABLE simulations: a feedback model for galaxies, groups, and clusters}}.
\newblock \emph{\bibinfo{journal}{\mnras}} \textbf{\bibinfo{volume}{479}}, \bibinfo{pages}{5385--5412} (\bibinfo{year}{2018}).

\bibitem{plan15}
\bibinfo{author}{{Planck Collaboration}} \emph{et~al.}
\newblock \bibinfo{title}{{Planck 2015 results. XIII. Cosmological parameters}}.
\newblock \emph{\bibinfo{journal}{\aap}} \textbf{\bibinfo{volume}{594}}, \bibinfo{pages}{A13} (\bibinfo{year}{2016}).

\bibitem{zhou26}
\bibinfo{author}{{Zhou}, D.} \emph{et~al.}
\newblock \bibinfo{title}{{Sunyaev-Zeldovich detection of hot intracluster gas at redshift 4.3}}.
\newblock \emph{\bibinfo{journal}{\nat}} \textbf{\bibinfo{volume}{649}}, \bibinfo{pages}{1130--1133} (\bibinfo{year}{2026}).

\bibitem{choi12}
\bibinfo{author}{{Choi}, E.}, \bibinfo{author}{{Ostriker}, J.~P.}, \bibinfo{author}{{Naab}, T.} \& \bibinfo{author}{{Johansson}, P.~H.}
\newblock \bibinfo{title}{{Radiative and Momentum-based Mechanical Active Galactic Nucleus Feedback in a Three-dimensional Galaxy Evolution Code}}.
\newblock \emph{\bibinfo{journal}{\apj}} \textbf{\bibinfo{volume}{754}}, \bibinfo{pages}{125} (\bibinfo{year}{2012}).

\bibitem{vayn21}
\bibinfo{author}{{Vayner}, A.} \emph{et~al.}
\newblock \bibinfo{title}{{Multiphase Outflows in High-redshift Quasar Host Galaxies}}.
\newblock \emph{\bibinfo{journal}{\apj}} \textbf{\bibinfo{volume}{923}}, \bibinfo{pages}{59} (\bibinfo{year}{2021}).

\bibitem{dima25}
\bibinfo{author}{{Di Mascolo}, L.} \emph{et~al.}
\newblock \bibinfo{title}{{Atacama Large Aperture Submillimeter Telescope (AtLAST) science: Resolving the hot and ionized Universe through the Sunyaev-Zeldovich effect}}.
\newblock \emph{\bibinfo{journal}{ORE}} \textbf{\bibinfo{volume}{4}}, \bibinfo{pages}{113} (\bibinfo{year}{2025}).

\bibitem{wang15}
\bibinfo{author}{{Wang}, L.} \emph{et~al.}
\newblock \bibinfo{title}{{Co-evolution of black hole growth and star formation from a cross-correlation analysis between quasars and the cosmic infrared background}}.
\newblock \emph{\bibinfo{journal}{\mnras}} \textbf{\bibinfo{volume}{449}}, \bibinfo{pages}{4476--4493} (\bibinfo{year}{2015}).

\bibitem{hall18}
\bibinfo{author}{{Hall}, K.~R.}, \bibinfo{author}{{Crichton}, D.}, \bibinfo{author}{{Marriage}, T.}, \bibinfo{author}{{Zakamska}, N.~L.} \& \bibinfo{author}{{Mandelbaum}, R.}
\newblock \bibinfo{title}{{Downsizing of star formation measured from the clustered infrared background correlated with quasars}}.
\newblock \emph{\bibinfo{journal}{\mnras}} \textbf{\bibinfo{volume}{480}}, \bibinfo{pages}{149--181} (\bibinfo{year}{2018}).

\bibitem{riek09}
\bibinfo{author}{{Rieke}, G.~H.} \emph{et~al.}
\newblock \bibinfo{title}{{Determining Star Formation Rates for Infrared Galaxies}}.
\newblock \emph{\bibinfo{journal}{\apj}} \textbf{\bibinfo{volume}{692}}, \bibinfo{pages}{556--573} (\bibinfo{year}{2009}).

\bibitem{fore13}
\bibinfo{author}{{Foreman-Mackey}, D.}, \bibinfo{author}{{Hogg}, D.~W.}, \bibinfo{author}{{Lang}, D.} \& \bibinfo{author}{{Goodman}, J.}
\newblock \bibinfo{title}{{emcee: The MCMC Hammer}}.
\newblock \emph{\bibinfo{journal}{\pasp}} \textbf{\bibinfo{volume}{125}}, \bibinfo{pages}{306} (\bibinfo{year}{2013}).

\bibitem{dezo10}
\bibinfo{author}{{de Zotti}, G.}, \bibinfo{author}{{Massardi}, M.}, \bibinfo{author}{{Negrello}, M.} \& \bibinfo{author}{{Wall}, J.}
\newblock \bibinfo{title}{{Radio and millimeter continuum surveys and their astrophysical implications}}.
\newblock \emph{\bibinfo{journal}{Astron. Astrophys. Rev.}} \textbf{\bibinfo{volume}{18}}, \bibinfo{pages}{1--65} (\bibinfo{year}{2010}).

\bibitem{kenn98}
\bibinfo{author}{{Kennicutt}, R.~C., Jr.}
\newblock \bibinfo{title}{{The Global Schmidt Law in Star-forming Galaxies}}.
\newblock \emph{\bibinfo{journal}{\apj}} \textbf{\bibinfo{volume}{498}}, \bibinfo{pages}{541--552} (\bibinfo{year}{1998}).

\bibitem{kirk15}
\bibinfo{author}{{Kirkpatrick}, A.} \emph{et~al.}
\newblock \bibinfo{title}{{The Role of Star Formation and an AGN in Dust Heating of z = 0.3-2.8 Galaxies. I. Evolution with Redshift and Luminosity}}.
\newblock \emph{\bibinfo{journal}{\apj}} \textbf{\bibinfo{volume}{814}}, \bibinfo{pages}{9} (\bibinfo{year}{2015}).

\bibitem{tazz18}
\bibinfo{author}{{Tazzari}, M.}, \bibinfo{author}{{Beaujean}, F.} \& \bibinfo{author}{{Testi}, L.}
\newblock \bibinfo{title}{{GALARIO: a GPU accelerated library for analysing radio interferometer observations}}.
\newblock \emph{\bibinfo{journal}{\mnras}} \textbf{\bibinfo{volume}{476}}, \bibinfo{pages}{4527--4542} (\bibinfo{year}{2018}).

\bibitem{tazz21}
\bibinfo{author}{Tazzari, M.}, \bibinfo{author}{Cronin-Coltsmann, P.} \& \bibinfo{author}{Hendler, N.}
\newblock \bibinfo{title}{mtazzari/uvplot: 0.2.11} 
\newblock Zenodo, /10.5281/zenodo.4574577 (\bibinfo{year}{2021}).

\bibitem{benn24}
\bibinfo{author}{{Bennett}, J.~S.}, \bibinfo{author}{{Sijacki}, D.}, \bibinfo{author}{{Costa}, T.}, \bibinfo{author}{{Laporte}, N.} \& \bibinfo{author}{{Witten}, C.}
\newblock \bibinfo{title}{{The growth of the gargantuan black holes powering high-redshift quasars and their impact on the formation of early galaxies and protoclusters}}.
\newblock \emph{\bibinfo{journal}{\mnras}} \textbf{\bibinfo{volume}{527}}, \bibinfo{pages}{1033--1054} (\bibinfo{year}{2024}).

\bibitem{rowe11}
\bibinfo{author}{{Rowe}, B.} \& \bibinfo{author}{{Silk}, J.}
\newblock \bibinfo{title}{{The Sunyaev-Zel'dovich effect due to hyperstarburst galaxy winds}}.
\newblock \emph{\bibinfo{journal}{\mnras}} \textbf{\bibinfo{volume}{412}}, \bibinfo{pages}{905--910} (\bibinfo{year}{2011}).

\bibitem{spri10}
\bibinfo{author}{{Springel}, V.}
\newblock \bibinfo{title}{{E pur si muove: Galilean-invariant cosmological hydrodynamical simulations on a moving mesh}}.
\newblock \emph{\bibinfo{journal}{\mnras}} \textbf{\bibinfo{volume}{401}}, \bibinfo{pages}{791--851} (\bibinfo{year}{2010}).

\bibitem{angu12}
\bibinfo{author}{{Angulo}, R.~E.} \emph{et~al.}
\newblock \bibinfo{title}{{Scaling relations for galaxy clusters in the Millennium-XXL simulation}}.
\newblock \emph{\bibinfo{journal}{\mnras}} \textbf{\bibinfo{volume}{426}}, \bibinfo{pages}{2046--2062} (\bibinfo{year}{2012}).

\bibitem{benn20}
\bibinfo{author}{{Bennett}, J.~S.} \& \bibinfo{author}{{Sijacki}, D.}
\newblock \bibinfo{title}{{Resolving shocks and filaments in galaxy formation simulations: effects on gas properties and star formation in the circumgalactic medium}}.
\newblock \emph{\bibinfo{journal}{\mnras}} \textbf{\bibinfo{volume}{499}}, \bibinfo{pages}{597--615} (\bibinfo{year}{2020}).

\bibitem{voge13}
\bibinfo{author}{{Vogelsberger}, M.} \emph{et~al.}
\newblock \bibinfo{title}{{A model for cosmological simulations of galaxy formation physics}}.
\newblock \emph{\bibinfo{journal}{\mnras}} \textbf{\bibinfo{volume}{436}}, \bibinfo{pages}{3031--3067} (\bibinfo{year}{2013}).

\bibitem{sija07}
\bibinfo{author}{{Sijacki}, D.}, \bibinfo{author}{{Springel}, V.}, \bibinfo{author}{{Di Matteo}, T.} \& \bibinfo{author}{{Hernquist}, L.}
\newblock \bibinfo{title}{{A unified model for AGN feedback in cosmological simulations of structure formation}}.
\newblock \emph{\bibinfo{journal}{\mnras}} \textbf{\bibinfo{volume}{380}}, \bibinfo{pages}{877--900} (\bibinfo{year}{2007}).

\end{thebibliography}

\begin{thebibliography}{10}
\expandafter\ifx\csname url\endcsname\relax
  \def\url#1{\burl{#1}}\fi
\expandafter\ifx\csname urlprefix\endcsname\relax\def\urlprefix{URL }\fi
\providecommand{\bibinfo}[2]{#2}
\providecommand{\eprint}[2][]{\url{#2}}
\providecommand{\doi}[1]{\url{https://doi.org/#1}}

\bibitem{beel06}
\bibinfo{author}{{Beelen}, A.} \emph{et~al.}
\newblock \bibinfo{title}{{350 {\ensuremath{\mu}}m Dust Emission from High-Redshift Quasars}}.
\newblock \emph{\bibinfo{journal}{\apj}} \textbf{\bibinfo{volume}{642}}, \bibinfo{pages}{694--701} (\bibinfo{year}{2006}).

\bibitem{kirk12}
\bibinfo{author}{{Kirkpatrick}, A.} \emph{et~al.}
\newblock \bibinfo{title}{{GOODS-Herschel: Impact of Active Galactic Nuclei and Star Formation Activity on Infrared Spectral Energy Distributions at High Redshift}}.
\newblock \emph{\bibinfo{journal}{\apj}} \textbf{\bibinfo{volume}{759}}, \bibinfo{pages}{139} (\bibinfo{year}{2012}).

\bibitem{bend25}
\bibinfo{author}{{Bendo}, G.~J.} \emph{et~al.}
\newblock \bibinfo{title}{{The dust emissivity index {\ensuremath{\beta}} in infrared-bright galaxies at $1.5 < z < 4.2$}}.
\newblock \emph{\bibinfo{journal}{\mnras}} \textbf{\bibinfo{volume}{540}}, \bibinfo{pages}{1560--1577} (\bibinfo{year}{2025}).

\bibitem{riek09}
\bibinfo{author}{{Rieke}, G.~H.} \emph{et~al.}
\newblock \bibinfo{title}{{Determining Star Formation Rates for Infrared Galaxies}}.
\newblock \emph{\bibinfo{journal}{\apj}} \textbf{\bibinfo{volume}{692}}, \bibinfo{pages}{556--573} (\bibinfo{year}{2009}).

\bibitem{bell03}
\bibinfo{author}{{Bell}, E.~F.}
\newblock \bibinfo{title}{{Estimating Star Formation Rates from Infrared and Radio Luminosities: The Origin of the Radio-Infrared Correlation}}.
\newblock \emph{\bibinfo{journal}{\apj}} \textbf{\bibinfo{volume}{586}}, \bibinfo{pages}{794--813} (\bibinfo{year}{2003}).

\bibitem{kirk19}
\bibinfo{author}{{Kirkpatrick}, A.}, \bibinfo{author}{{Sharon}, C.}, \bibinfo{author}{{Keller}, E.} \& \bibinfo{author}{{Pope}, A.}
\newblock \bibinfo{title}{{CO Emission in Infrared-selected Active Galactic Nuclei}}.
\newblock \emph{\bibinfo{journal}{\apj}} \textbf{\bibinfo{volume}{879}}, \bibinfo{pages}{41} (\bibinfo{year}{2019}).

\bibitem{bisc21}
\bibinfo{author}{{Bischetti}, M.} \emph{et~al.}
\newblock \bibinfo{title}{{The WISSH quasars project. IX. Cold gas content and environment of luminous QSOs at z {\ensuremath{\sim}} 2.4-4.7}}.
\newblock \emph{\bibinfo{journal}{\aap}} \textbf{\bibinfo{volume}{645}}, \bibinfo{pages}{A33} (\bibinfo{year}{2021}).

\bibitem{cari13}
\bibinfo{author}{{Carilli}, C.~L.} \& \bibinfo{author}{{Walter}, F.}
\newblock \bibinfo{title}{{Cool Gas in High-Redshift Galaxies}}.
\newblock \emph{\bibinfo{journal}{\araa}} \textbf{\bibinfo{volume}{51}}, \bibinfo{pages}{105--161} (\bibinfo{year}{2013}).

\bibitem{bisc17}
\bibinfo{author}{{Bischetti}, M.} \emph{et~al.}
\newblock \bibinfo{title}{{The WISSH quasars project. I. Powerful ionised outflows in hyper-luminous quasars}}.
\newblock \emph{\bibinfo{journal}{\aap}} \textbf{\bibinfo{volume}{598}}, \bibinfo{pages}{A122} (\bibinfo{year}{2017}).

\bibitem{pern18}
\bibinfo{author}{{Perna}, M.} \emph{et~al.}
\newblock \bibinfo{title}{{Molecular gas content in obscured AGN at $z > 1$}}.
\newblock \emph{\bibinfo{journal}{\aap}} \textbf{\bibinfo{volume}{619}}, \bibinfo{pages}{A90} (\bibinfo{year}{2018}).

\bibitem{sarg14}
\bibinfo{author}{{Sargent}, M.~T.} \emph{et~al.}
\newblock \bibinfo{title}{{Regularity Underlying Complexity: A Redshift-independent Description of the Continuous Variation of Galaxy-scale Molecular Gas Properties in the Mass-star Formation Rate Plane}}.
\newblock \emph{\bibinfo{journal}{\apj}} \textbf{\bibinfo{volume}{793}}, \bibinfo{pages}{19} (\bibinfo{year}{2014}).

\bibitem{zaka14}
\bibinfo{author}{{Zakamska}, N.~L.} \& \bibinfo{author}{{Greene}, J.~E.}
\newblock \bibinfo{title}{{Quasar feedback and the origin of radio emission in radio-quiet quasars}}.
\newblock \emph{\bibinfo{journal}{\mnras}} \textbf{\bibinfo{volume}{442}}, \bibinfo{pages}{784--804} (\bibinfo{year}{2014}).

\bibitem{pane19}
\bibinfo{author}{{Panessa}, F.} \emph{et~al.}
\newblock \bibinfo{title}{{The origin of radio emission from radio-quiet active galactic nuclei}}.
\newblock \emph{\bibinfo{journal}{Nature Astronomy}} \textbf{\bibinfo{volume}{3}}, \bibinfo{pages}{387--396} (\bibinfo{year}{2019}).

\bibitem{alho22}
\bibinfo{author}{{Alhosani}, A.} \emph{et~al.}
\newblock \bibinfo{title}{{Dependence of the Radio Emission on the Eddington Ratio of Radio-quiet Quasars}}.
\newblock \emph{\bibinfo{journal}{\apj}} \textbf{\bibinfo{volume}{936}}, \bibinfo{pages}{73} (\bibinfo{year}{2022}).

\bibitem{chen25}
\bibinfo{author}{{Chen}, S.} \emph{et~al.}
\newblock \bibinfo{title}{{A Dichotomy in the 1{\textendash}24 GHz Parsec-scale Radio Spectra of Radio-quiet Quasars}}.
\newblock \emph{\bibinfo{journal}{\apj}} \textbf{\bibinfo{volume}{979}}, \bibinfo{pages}{241} (\bibinfo{year}{2025}).

\bibitem{bald22}
\bibinfo{author}{{Baldi}, R.~D.} \emph{et~al.}
\newblock \bibinfo{title}{{The PG-RQS survey. Building the radio spectral distribution of radio-quiet quasars. I. The 45-GHz data}}.
\newblock \emph{\bibinfo{journal}{\mnras}} \textbf{\bibinfo{volume}{510}}, \bibinfo{pages}{1043--1058} (\bibinfo{year}{2022}).

\bibitem{klei18}
\bibinfo{author}{{Klein}, U.}, \bibinfo{author}{{Lisenfeld}, U.} \& \bibinfo{author}{{Verley}, S.}
\newblock \bibinfo{title}{{Radio synchrotron spectra of star-forming galaxies}}.
\newblock \emph{\bibinfo{journal}{\aap}} \textbf{\bibinfo{volume}{611}}, \bibinfo{pages}{A55} (\bibinfo{year}{2018}).

\bibitem{poll00}
\bibinfo{author}{{Polletta}, M.}, \bibinfo{author}{{Courvoisier}, T.~J.-L.}, \bibinfo{author}{{Hooper}, E.~J.} \& \bibinfo{author}{{Wilkes}, B.~J.}
\newblock \bibinfo{title}{{The far-infrared emission of radio loud and radio quiet quasars}}.
\newblock \emph{\bibinfo{journal}{\aap}} \textbf{\bibinfo{volume}{362}}, \bibinfo{pages}{75--96} (\bibinfo{year}{2000}).

\bibitem{azad23}
\bibinfo{author}{{Azadi}, M.} \emph{et~al.}
\newblock \bibinfo{title}{{Disentangling the AGN and Star formation Contributions to the Radio-X-Ray Emission of Radio-loud Quasars at $1 < Z < 2$}}.
\newblock \emph{\bibinfo{journal}{\apj}} \textbf{\bibinfo{volume}{945}}, \bibinfo{pages}{145} (\bibinfo{year}{2023}).

\bibitem{spea20}
\bibinfo{author}{{Speagle}, J.~S.}
\newblock \bibinfo{title}{{DYNESTY: a dynamic nested sampling package for estimating Bayesian posteriors and evidences}}.
\newblock \emph{\bibinfo{journal}{\mnras}} \textbf{\bibinfo{volume}{493}}, \bibinfo{pages}{3132--3158} (\bibinfo{year}{2020}).

\bibitem{kopo25}
\bibinfo{author}{Koposov, S.} \emph{et~al.}
\newblock \bibinfo{title}{joshspeagle/dynesty: v3.0.0} (\bibinfo{year}{2025}).
\newblock \urlprefix\url{https://doi.org/10.5281/zenodo.17268284}.

\bibitem{higs19}
\bibinfo{author}{{Higson}, E.}, \bibinfo{author}{{Handley}, W.}, \bibinfo{author}{{Hobson}, M.} \& \bibinfo{author}{{Lasenby}, A.}
\newblock \bibinfo{title}{{Dynamic nested sampling: an improved algorithm for parameter estimation and evidence calculation}}.
\newblock \emph{\bibinfo{journal}{Statistics and Computing}} \textbf{\bibinfo{volume}{29}}, \bibinfo{pages}{891--913} (\bibinfo{year}{2019}).

\bibitem{fero09}
\bibinfo{author}{{Feroz}, F.}, \bibinfo{author}{{Hobson}, M.~P.} \& \bibinfo{author}{{Bridges}, M.}
\newblock \bibinfo{title}{{MULTINEST: an efficient and robust Bayesian inference tool for cosmology and particle physics}}.
\newblock \emph{\bibinfo{journal}{\mnras}} \textbf{\bibinfo{volume}{398}}, \bibinfo{pages}{1601--1614} (\bibinfo{year}{2009}).

\bibitem{neal03}
\bibinfo{author}{Neal, R.~M.}
\newblock \bibinfo{title}{{Slice sampling}}.
\newblock \emph{\bibinfo{journal}{The Annals of Statistics}} \textbf{\bibinfo{volume}{31}}, \bibinfo{pages}{705 -- 767} (\bibinfo{year}{2003}).

\bibitem{hand15a}
\bibinfo{author}{{Handley}, W.~J.}, \bibinfo{author}{{Hobson}, M.~P.} \& \bibinfo{author}{{Lasenby}, A.~N.}
\newblock \bibinfo{title}{{polychord: nested sampling for cosmology.}}
\newblock \emph{\bibinfo{journal}{\mnras}} \textbf{\bibinfo{volume}{450}}, \bibinfo{pages}{L61--L65} (\bibinfo{year}{2015}).

\bibitem{hand15b}
\bibinfo{author}{{Handley}, W.~J.}, \bibinfo{author}{{Hobson}, M.~P.} \& \bibinfo{author}{{Lasenby}, A.~N.}
\newblock \bibinfo{title}{{POLYCHORD: next-generation nested sampling}}.
\newblock \emph{\bibinfo{journal}{\mnras}} \textbf{\bibinfo{volume}{453}}, \bibinfo{pages}{4384--4398} (\bibinfo{year}{2015}).

\bibitem{ragi16}
\bibinfo{author}{{Raginski}, I.} \& \bibinfo{author}{{Laor}, A.}
\newblock \bibinfo{title}{{AGN coronal emission models - I. The predicted radio emission}}.
\newblock \emph{\bibinfo{journal}{\mnras}} \textbf{\bibinfo{volume}{459}}, \bibinfo{pages}{2082--2096} (\bibinfo{year}{2016}).

\end{thebibliography}

\end{document}